\documentstyle[preprint,aps]{revtex}

\begin{document}

\preprint{
\font\fortssbx=cmssbx10 scaled \magstep2
\hbox to \hsize{
\hskip.5in \raise.1in\hbox{\fortssbx University of Wisconsin -
Madison}
\hfill$\vtop{\hbox{\bf MADPH-95-870}
                \hbox{\bf IUHET-294}
                \hbox{\bf RAL-95-013}
                \hbox{\bf hep-ph/9503204}
                \hbox{March 1995}}$ }
}

\title{\vspace*{.1in}
Quark Singlets: Implications and Constraints }

\author{V. Barger$^a$, M.S.~Berger$^b$
and R.J.N. Phillips$^c$}

\address{
$^a$Physics Department, University of Wisconsin, Madison, WI 53706,
USA\\
$^b$Physics Department, Indiana University, Bloomington, IN 47405,
USA\\
$^c$Rutherford Appleton Laboratory, Chilton, Didcot, Oxon OX11 0QX,
UK}

\maketitle

\thispagestyle{empty}

\begin{abstract}
Quarks whose left- and right-handed chiral components are both singlets
with respect to the SU(2) weak-isospin gauge group, offer interesting
physics possibilities beyond the Standard Model (SM) already studied
in many contexts. We here address some further aspects. We first collect
and update the constraints from present data on their masses and mixings
with conventional quarks. We discuss possible effects on $b\to s\gamma$
and $Z\to b\bar b$ decays and give fresh illustrations of CP asymmetries
in $B^0$ decays differing dramatically from SM expectations. We analyse
singlet effects in grand unification scenarios: $d$-type singlets are
most economically introduced in ${\bf 5+5^*}$ multiplets of $SU(5)$,
with up to three generations, preserving gauge coupling unification with
perturbative values up to the GUT scale; $u$-type singlets can arise in
${\bf 10+10^*}$ multiplets of $SU(5)$ with at most one light generation.
With extra matter multiplets the gauge couplings are bigger; we give the
two-loop evolution equations including exotic multiplets and a possible
extra $U(1)$ symmetry. Two-loop effects can become important, threatening
unification (modulo threshold effects), perturbativity and asymptotic
freedom of $\alpha_3$. In the Yukawa sector, top-quark fixed-point
behaviour is preserved and singlet-quark couplings have infrared fixed
points too, but unification of $b$ and $\tau$ couplings is not possible
in a three-generation $E_6$ model.
\end{abstract}

\newpage

\section{Introduction}

In addition to the established three generations of quarks in the
Standard Model (SM), the possible existence of exotic singlet
quarks (whose left and right chiral components are both singlets
with respect to the SU(2) weak isospin gauge group)
has been raised in various contexts.   It was once
questioned whether the $b$ quark might be such a singlet, with
no doublet partner $t$\cite{bsing}.  One
charge $-{1\over 3}$ singlet quark appears naturally in each
27-plet fermion generation of $E_6$ Grand Unification Theories (GUTs)
\cite{rosner,bdpw,rizzo}.  Charge ${2\over 3}$ singlet quarks have
been variously motivated, as part of a new mass mechanism
for top quarks \cite{bh} or as part of a new supersymmetric gauge
model with natural baryon-number conservation\cite{ma}.  If they
exist, both kinds of singlet quarks can be produced via their strong
and electroweak gauge couplings; mixing with standard quarks then
allows the mixed mass eigenstates to decay via charged currents (CC)
or neutral currents (NC) to lighter quarks $q$ plus $W$ or $Z$
\cite{bdpw,rizzo,alz}, and also via Yukawa couplings to $q$ plus
Higgs bosons $H$ \cite{akq,bw}.   Singlet quark production and
decay can therefore give characteristic new signals and
modifications of old signals, discussed in the literature
\cite{bdpw,rizzo,bh,alz,akq,bp,agra,mukho,dp,hou,bm,bp2}.
Possible indirect consequences of singlet-quark mixing for
flavor-changing neutral currents (FCNC),  flavor-diagonal neutral
currents (FDNC)  and CP violation have also been considered~
\cite{bh,acc,branco1,mukho2,ll,nr,nir,branco2,branco3,cs,branco4}

In the present paper we address some further aspects of singlet quark
physics.   We first collect and update the direct and indirect
constraints on masses and singlet-doublet mixing from present data,
illustrating possible effects on $b\to s\gamma$ and $Z\to b\bar b$
decays and on CP asymmetries in neutral $B$ decays.  We then analyse
the impact of $Q=-{1\over 3}$ and $Q={2\over 3}$ singlet quarks on
the renormalization group equations (RGE), on the unification and
perturbativity of gauge and Yukawa couplings, and on the exotic matter
multiplets in GUT scenarios.

Section II introduces our notation and lists general basic properties
of singlet quark couplings and mixings with SM quarks.   Section III
addresses the $4\times 4$ mixing matrix, arising when one singlet
mixes with three SM quarks, and extracts the full set of unitarity
constraints based on present limits on the CKM submatrix.  Section IV
discusses the constraints implied by the absence of identifiable signals
from singlet-quark production and decay at present $e^+e^-$ and
$p\bar p$ colliders.  Section V considers tree and box diagram
contributions to neutral meson-antimeson oscillations and the indirect
constraints on singlet-quark mixing from present data.
Section VI addresses indirect constraints from FCNC and
FDNC decays, including a new more stringent measurement of $K_L\to\mu^+
\mu^-$ and weak bounds from $B^0,D^0\to\mu^+\mu^-$ limits; the topical
cases $b\to s\gamma$ and $Z\to b\bar b$ are discussed here.  The
global FDNC constraints are comprehensive enough to have useful
repercussions via unitarity, for $d$-type singlet mixing.  Section VII
discusses CP asymmetries in neutral $B$ decays, with new illustrations
of how $d$-type singlet mixing can give dramatic changes from SM
expectations.  Section VIII, our major new contribution, analyses
the possible roles of singlet quarks in GUT scenarios.  We show that
$d$-type singlets are most economically introduced in ${\bf 5\, + \,5^*}$
multiplets of $SU(5)$, with up to three generations, preserving gauge
coupling unification and perturbativity up to the GUT scale; $u$-type
singlets can arise in ${\bf 10\, +\, 10^*}$ multiplets of $SU(5)$ with
at most one light generation.  The presence of extra matter multiplets
makes the gauge couplings bigger and two-loop effects potentially more
important. We give the two-loop evolution equations, including the effects
of exotic matter multiplets and a possible additional $U(1)'$ gauge
coupling, and show that two-loop effects can threaten not only unification
(where threshold effects may partly compensate) but also perturbativity
and asymptotic freedom of $\alpha_3$ at large scales.  In the Yukawa
sector, top-quark fixed-point behaviour is preserved and singlet-quark
couplings have infrared fixed points too, but unification of $b$ and
$\tau$ couplings is not possible in a three-generation $E_6$ model.
Finally, Section IX summarizes our conclusions while Appendices A and B
contain some technical details.

\section{Basic properties and notation}

We shall generally denote singlet quarks by the symbol $x$, and
SM quarks by $q$.  More specifically, $x_d$ denotes
a generic charge $-{1\over 3}$ singlet and $x_u$ implies
charge ${2\over 3}$.  The weak isospin $T_3$ and hypercharge
${1\over 2}Y$ of the left and right chiral components,
characterizing their SU(2) and U(1) gauge couplings, contrast
with SM assignments as follows:
\[ \arraycolsep=1em
\begin{array}{ccccccc}
           &   u_L    &   d_L     &   x_{uL} &  x_{dL}
                                                    &u_R,x_{uR}&d_R,x_{dR} \\
  T_3      &{1\over 2}&-{1\over 2}&    0     &  \ 0
                                                   &    0     &  \  0       \\
{1\over 2}Y&{1\over 6}&\ {1\over 6}&{2\over 3}&-{1\over 3}
                                                   &{2\over  3}&-{1\over 3}\\
   Q       &{2\over 3}&-{1\over 3}&{2\over 3}&-{1\over 3}
                                                   &{2\over  3}&-{1\over 3}
\end{array}
\]
where $Q=T_3 + {1\over 2}Y$ is the electric charge.
The vector and axial couplings to $Z$ are
\[ \arraycolsep=1em
\begin{array}{ccccc}
 & u & d & x_u & x_d \\
g_V & {1\over4}-{2\over3}\sin^2\theta_W &
-{1\over4}+{1\over3}\sin^2\theta_W &
-{2\over3}\sin^2\theta_W & {1\over3}\sin^2\theta_W \\
g_A & -{1\over4} & {1\over4} & &
\end{array}
\]
where $\theta_W$ is the Weinberg angle.  Both SM and singlet quarks
are color triplets and have the same couplings to gluons $g$. Hence
singlet quarks have pure vector gauge couplings to $g, \gamma, Z$
(and zero coupling to $W$); they are sometimes called ``vector-like"
or more precisely ``vector-singlet" quarks.  They do not contribute
to chiral anomalies.

\medskip
\noindent\underline{(a) $2\times2$ quark mixing example}

Yukawa interactions with Higgs fields generate quark masses and
mixings.   Mixing with conventional quarks provides natural decay
channels and is expected at some level, since new quarks are
necessarily unstable\cite{nr2}.  Suppose first, for simplicity,
that mass eigenstates $q,x$ arise from the mixing of just one SM
quark field $q'$ with a singlet quark field $x'$ of the same
(unspecified) charge. Then the SM Higgs field $H$ can generate a
$m'\bar q'_Lx'_R+$h.c. mixing term as well as the usual
$m\bar q'_Lq'_R+$h.c. mass term.   A pure singlet mass term
$M\bar x'_L x'_R+{}$h.c.\  requires an isosinglet Higgs field
$S$ with vacuum expectation value $v_S$ and coupling
$(M/v_S)S\bar x'_Lx'_R+{}$h.c.; this field can also generate a
$M'\bar x'_Lq'_R$ term.  We then have the $2\times2$ mass matrix
\begin{equation}
\arraycolsep=1em
\left(\begin{array}{cc}
m & m'\\
M' & M
\end{array}\right)
\end{equation}
where the rows refer to $\bar q'_L,\ \bar x'_L$ and the columns
refer to $q'_R,x'_R$. This is diagonalized by independent rotations
of $L$ and $R$ coordinates, giving quark mass eigenstates
$q$ and $x$:
\begin{eqnarray}
q_L &=& q'_L\cos\theta_L-x'_L\sin\theta_L\;, \qquad
q_R = q'_R\cos\theta_R-x'_R\sin\theta_R\;,\\
x_L &=& q'_L\sin\theta_L+x'_L\cos\theta_L\;, \qquad
x_R = q'_R\sin\theta_R+x'_R\cos\theta_R\;.
\end{eqnarray}
Since no singlets have yet been discovered, it is natural to assume
that the mixing angles $\theta_L,\ \theta_R$ are small and $x$ is
much heavier than $q$ (at least for $q=u,d,s,c,b$), with
$m_q\simeq m$, $m_x\simeq M\gg m, m', M'$.  Then $q$ and $x$ are
dominated by $q'$ and $x'$ components,  respectively, with
$\theta_L\simeq m'/M,\ \theta_R\simeq M'/M$.  Note that  $SU(2)_L$
gauge couplings relate exclusively to $q'_L$ and hence are
controlled by the left-handed mixing angle $\theta_L$ only.

The heavy mostly-singlet quark $x$ can now decay to $q''W$ and $qZ$
via the couplings
\begin{eqnarray}
{\cal L}_{xq''W} & = & -{g\over\sqrt 2} \sin\theta_L \bar q''_L
\gamma^\mu W_\mu x_L \; , \\
{\cal L}_{xqZ}   & = & -{g_Z\over 2} \sin\theta_L \cos\theta_L
\bar q_L \gamma^\mu Z_\mu x_L \; .
\end{eqnarray}
where $g$ is the SU(2) gauge coupling, $g_Z=g/\cos\theta_W$ and
$q''_L$ is the combination of light quarks that couple via $W$
to $q'_L$.  Since $\cos^2\theta_L\simeq 1$ by assumption, this
gives branching fractions in the ratio  $B(x\to q''W)/B(x\to qZ)
\simeq 2$ up to phase space  factors~\cite{bdpw}.  Furthermore,
if the SM Higgs boson is light enough, $x$ can also decay to
$qH$ via the coupling
\begin{equation}
 {\cal L}_{xqH} = -{gm'\over 2M_W} \bar q_L H x_R
\simeq -{g\sin\theta_L m_X\over 2M_W} \bar q_L H x_R \;.
\end{equation}
Hence the Higgs decay mode too is scaled by $\sin\theta_L$, and the
three decay branching fractions are in the ratios~\cite{akq}
\begin{equation}
B(x\to q''W) : B(x\to qZ) : B(x\to qH) \simeq 2 : 1 : 1
\end{equation}
up to phase space factors that are close to 1, if $m_x\gg M_W, M_Z,
M_H, m_q, m_{q''}$. These ratios can however be altered greatly if
this mass ordering does not hold, or if there is large mixing
{}~\cite{akq,mukho,dp,hou,bp2}.

In general singlet quarks can mix with all SM quarks of the same
charge, requiring a more extended formalism. We first consider
scenarios with just one new singlet quark.

\medskip
\noindent\underline{(b) One $Q=-{1\over3}$ singlet quark mixing}

For the case of one charge $-{1\over3}$ singlet field, mixing with
the three SM fields of this charge, we denote the mass eigenstate
by $d,s,b,x$ where the first three are identified with the known
quarks (now carrying hitherto unsuspected singlet components) and
$x$ is still undiscovered. We denote by $d'_L, s'_L, b'_L$ the
three orthonormal linear combinations of left chiral components
that are SU(2)$_L$ doublet partners of the known $Q={2\over3}$
fields $u_L,c_L,t_L$; the remaining orthonormal combination
$x'_L$ is an SU(2)$_L$ singlet, and we can write
\begin{equation}
\left(\begin{array}{c}
d'_L\\
s'_L\\
b'_L\\
x'_L
\end{array}\right) =
\left(\begin{array}{cccc}
V_{ud} & V_{us} & V_{ub} & V_{ux} \\
V_{cd} & V_{cs} & V_{cb} & V_{cx} \\
V_{td} & V_{ts} & V_{tb} & V_{tx} \\
V_{od} & V_{os} & V_{ob} & V_{ox}
\end{array}\right)
\left(\begin{array}{c}
d_L\\
s_L\\
b_L\\
x_L
\end{array}\right)  \;.
\label{eq:defv}
\end{equation}
Here the $4\times4$ unitary matrix $V$ generalizes the
Cabibbo-Kobayashi-Maskawa (CKM) matrix $V_{\rm CKM}$. The top three
rows of $V$ control the SU(2)$_L$ gauge couplings of $W$ and $Z$
bosons; the first 3 rows and columns of $V$ are precisely
$V_{\rm CKM}$. The submatrix $V_{\rm CKM}$ is generally non-unitary.

The $Z$ couplings to the SU(2)$_L$ left-handed doublet and singlet
weak eigenstates $q'_L=d'_L, s'_L, b'_L, x'_L$ are given by
\begin{equation}
{\cal L} = -g_Z \sum_{q'} \bar q'_L \left (T_3 +{1\over 3}\sin^2\theta_W
\right )
\gamma^\mu Z_\mu q'_L \; .
\end{equation}
Hence the FCNC couplings between the mass eigenstates
$q_i = d,s,b,x$ are
\begin{eqnarray}
{\cal L}_{\rm FCNC} &=& \case1/2 g_Z \sum_{i\neq j} z_{ij} \bar
q_{iL}
\gamma^\mu Z_\mu q_{jL} \;, \label{eq:Lfcnc}\\
z_{ij} &=& V^*_{ui} V_{uj} + V^*_{ci} V_{cj} + V^*_{ti} V_{tj} =
\delta_{ij}
-V^*_{oi} V_{oj} \; ,
\label{eq:dzij}
\end{eqnarray}
using the unitarity of $V$.  Thus the FCNC coefficients $z_{ij}$
are measures of non-unitarity in $V_{CKM}$.
The corresponding FDNC couplings are
\begin{equation}
{\cal L}_{\rm FDNC} = g_Z \sum_{i=d,s,b,x} \bar q_i\gamma^\mu Z_\mu
\left [{1\over 4}z_{ii}(1-\gamma_5)-{1\over 3}\sin^2\theta_W\right ]q_i \;.
\label{eq:dfdnc}
\end{equation}
Thus for the standard $d,s,b$ quarks, mixing with $x$ reduces direct
left-handed FDNC by a factor $(z_{ii}-{2\over 3}\sin^2\theta_W) /
(1-{2\over 3}\sin^2\theta_W)$ and leaves right-handed FDNC unchanged.

Hence for $m_x > M_W, M_Z$, the tree-level widths for CC and FCNC
decays to light quarks are
\begin{eqnarray}
\Gamma(x\to q_iW) &=& {G_F m_x^3 \over 8\pi\sqrt 2}
   \left (1 - {M_W^2\over m_x^2}\right )^2
\left (1+{2M_W^2\over m_x^2}\right ) |V_{ix}|^2\;,
\label{eq:gccxd}\\
\Gamma(x\to q_jZ) &=& {G_F m_x^3 \over 16\pi\sqrt 2}
   \left (1 - {M_Z^2\over m_x^2}\right )^2
\left (1+{2M_Z^2\over m_x^2}\right ) |z_{jx}|^2\;.
\label{eq:gfcncxd}
\end{eqnarray}
If $x$ is heavy enough that all the $x\to q_iW$ and $x\to q_jZ$
  channels are open and all the phase space factors are $\simeq 1$,
  then we can use unitarity to sum over $i=u,c,t$ and $j=d,s,b$ and
  obtain the total CC and FCNC decay widths,
  \begin{eqnarray}
  \Gamma(CC)&\simeq &{G_Fm_x^3\over 8\pi \sqrt 2}\Sigma_i|V_{ix}|^2
            \simeq {G_Fm_x^3\over 8\pi \sqrt 2}(1-|V_{ox}|^2)\;, \\
  \Gamma(FCNC)&\simeq &{G_Fm_x^3\over 16\pi \sqrt 2}
                     \Sigma_j|V^*_{oj}|^2 |V_{ox}|^2
        \simeq {G_Fm_x^3\over 16\pi \sqrt 2}(1-|V_{ox}|^2)|V_{ox}|^2\;.
  \end{eqnarray}
  Hence for small mixing ($|V_{ox}|\simeq 1$) we obtain
  \begin{equation}
  \Gamma (CC)/\Gamma (FCNC) \simeq 2,
  \end{equation}
  a result proved earlier for two-quark mixing, modulo phase
  space factors.

\medskip
\noindent\underline{(c) One $Q={2\over3}$ singlet quark mixing}

Consider now one $Q={2\over3}$ singlet field  mixing with the SM
fields of the same charge and denote the mass eigenstates by
$u,c,t,x$,  identifying the first three with the known quarks.
Let $u'_L,c'_L,t'_L$ be the three orthonormal linear combinations
of left chiral components that form SU(2)$_L$ doublets with the
known $Q=-{1\over3}$ fields $d_L,s_L,b_L$,  respectively, while
the remaining combination $x'_L$ is a singlet. We can then write
\begin{equation}
\left(\begin{array}{cccc}\bar u'_L&\bar c'_L&\bar t'_L&\bar
x'_L\end{array}\right) =
\left(\begin{array}{cccc}\bar u_L&\bar c_L&\bar t_L&\bar
x_L\end{array}\right)
\left(\begin{array}{cccc}
\hat V_{ud} & \hat V_{us} & \hat V_{ub} & \hat V_{uo} \\
\hat V_{cd} & \hat V_{cs} & \hat V_{cb} & \hat V_{co} \\
\hat V_{td} & \hat V_{ts} & \hat V_{tb} & \hat V_{to} \\
\hat V_{xd} & \hat V_{xs} & \hat V_{xb} & \hat V_{xo}
\end{array}\right) \;.
\label{eq:defvhat}
\end{equation}
The first three rows of the unitary matrix $\hat V$ control the
SU(2)$_L$ couplings of $W$ and $Z$; the first three rows and
columns of $\hat V$ are precisely $V_{CKM}$ (now generally
non-unitary).  The $Z$ couplings to the SU(2)$_L$ left-handed
doublet and singlet weak eigenstates $q'_L =u'_L, c'_L, t'_L,
x'_L$ are given by
\begin{equation}
{\cal L} = g_Z \sum_{q'} \bar q'_L \left(-T_3+{2\over
3}\sin^2\theta_W
\right)\gamma^\mu Z_\mu q'_L
\label{eq:uzij}
\end{equation}
and the FCNC couplings between mass eigenstates $q_i = u,c,t,x$
are
\begin{equation}
{\cal L}_{\rm FCNC} = -\case1/2 g_Z \sum_{i\neq j} \hat z_{ij} \bar
q_{iL}
\gamma^\mu Z_\mu q_{jL} \;,
\label{eq:ufdnc}
\end{equation}
with
\begin{equation}
\hat z_{ij} = \hat V_{id} \hat V^*_{jd} + \hat V_{is} \hat V^*_{js} +
\hat
V_{ib} \hat V^*_{jb} = \delta_{ij} -\hat V_{io} \hat V^*_{jo} \;,
\end{equation}
using the unitarity of $\hat V$.  Here again the FCNC coefficients
$\hat z_{ij}$ are direct measures of non-unitarity in $V_{CKM}$.
The corresponding FDNC couplings are
\begin{equation}
{\cal L}_{\rm FDNC} = g_Z \sum_{i=u,c,t,x} \bar q_i\gamma^\mu Z_\mu
\left [-{1\over 4}\hat z_{ii}(1-\gamma_5)+{2\over 3}\sin^2\theta_W
\right ]q_i \;.
\end{equation}
For standard $u,c,t$ quarks, mixing with $x$ again reduces left-handed
FDNC and leaves right-handed FDNC unchanged.

The decay-width formulas are obtained from Eqs.(\ref{eq:gccxd})-
(\ref{eq:gfcncxd}), by substituting $\hat V_{xi}$ and $\hat z_{jx}$
for $V_{ix}$ and $z_{jx}$.

\medskip
\noindent\underline{(d) One $Q=-{1\over3}$ quark and one
$Q={2\over3}$ quark
mixing}

We here combine the notations of (b) and (c) above, and define
\begin{equation}
(\bar u'_L, \bar c'_L, \bar t'_L, \bar x'_{uL}) = (\bar u_L, \bar
c_L, \bar
t_L, \bar x_{uL}) \hat V
\end{equation}
to be three doublet and one singlet $Q={2\over3}$ fields, while
\begin{equation}
\left(\begin{array}{c}
d'_L\\
s'_L\\
b'_L\\
x'_{dL}
\end{array}\right)
= V\left(\begin{array}{c}
d_L\\
s_L\\
b_L\\
x_{dL}
\end{array}\right)
\end{equation}
are the corresponding three doublets (paired with $u'_L, c'_L, t'_L)$
and remaining singlet $Q=-\case1/3$ fields. Neutral-current couplings
of $x_d$ and $x_u$ are as in (b) and (c) above.   Charged-current
couplings are defined via the matrix $\hat V\Delta V$, where $\Delta$
is diagonal with elements 1, 1, 1, 0 down the diagonal; $\hat V\Delta
V$ generalizes the CKM matrix, its first three rows and columns being
simply $V_{\rm CKM}$. For more general mixing parametrizations, see
Refs.~\cite{branco1,ll}.

\section{$4\times4$ mixing matrix}

\medskip\noindent\underline{(a) Experimental constraints}

When extra quarks are mixed in, unitarity constraints no longer
apply to the $3\times3$ CKM submatrix.  Without these constraints,
the CKM matrix elements lie in the following ranges~\cite{gilman}:
\begin{equation}
|V| = \left(\begin{array}{cccc}
0.9728\; -\; 0.9757 & \quad 0.218\; -\; 0.224 & \quad 0.002\;
-\; 0.005 & \quad ..\\
0.180 \; -\; 0.228  & \quad 0.800\; -\; 0.975 & \quad 0.032\;
-\; 0.048 & \quad ..\\
0.0   \; -\; 0.013  & \quad 0.0  \; -\; 0.56  & \quad 0.0  \;
-\; 0.9995& \quad ..\\
  ..                 &           ..             & \quad ..
           & \quad ..
\end{array}\right)
\label{eq:ckmlim}
\end{equation}
However these numbers were obtained before the evidence for the top
quark at Fermilab~\cite{cdf}. The presence of an apparent top quark
signal in $b$-tagged events at, or even above, the predicted SM rate
\cite{cdf,dzero},  strongly suggests a dominant $t\to bW$ decay with
$|V_{tb}|\simeq1$. With this extra constraint, all the off-diagonal
elements of the $4\times4$ quark mixing matrix $V$ (or $\hat V$) are
necessarily small. One can then generalize the Wolfenstein
parameterization
\begin{equation}
V_{us}\sim\lambda\,, \quad V_{ub}\sim\lambda^3 A(\rho-i\eta)\,, \quad
V_{cb}\sim\lambda^2A\,, \quad V_{td}\sim\lambda^3A(1-\rho-i\eta) \,,
\end{equation}
by taking for example (for the $Q=-\case1/3$ case $V$)
\begin{equation}
V_{od}\sim B(\alpha-i\beta)\,, \quad V_{os}\sim B(\sigma-i\tau)\,,
\quad
V_{ob}\sim B \,.
\end{equation}
Here the new parameters $\alpha,\ \beta,\ \sigma,\ \tau,\ B$ are real
and $B$ is small (no hierarchy of these elements is imposed here).
Often it is more convenient to adopt a parameterization in
terms of the sines $s_i=\sin\theta_i$ of small angles $\theta_i$,
setting $\cos\theta_i\simeq1$ and neglecting all $s_is_j$ terms
except $s_1s_2$ (see also Ref.\cite{nir}):
\begin{equation}
V \simeq \left(\begin{array}{cccc}
1                       & s_1 & s_3e^{-i\delta_1} & s_6
\\
-s_1                    & 1   & s_2               & s_5e^{-i\delta_3}
\\
-s_3e^{i\delta_1}+s_1s_2&-s_2 & 1                 & s_4e^{-i\delta_2}
\\
-s_6   &-s_5e^{i\delta_3}     &-s_4e^{i\delta_2}  & 1
\end{array}\right) \;.
\end{equation}
In this parameterization, the FCNC coefficients ($z_{ij}=z_{ji}^*$)
are
\begin{equation}
\begin{array}{ccccccccc}
z_{ds}&=& -s_5s_6e^{i\delta_3},& z_{db}&=&-s_4s_6e^{i\delta_2},&
z_{sb}&=& -s_4s_5e^{i(\delta_2-\delta_3)}\\
z_{dx}&=&
s_6,&z_{sx}&=&s_5e^{-i\delta_3},&z_{bx}&=&s_4e^{-i\delta_2}.
\end{array}
\end{equation}
Similarly, if this parameterization is applied to $\hat V$ in the
case of one $Q={2\over 3}$ singlet quark, we have FCNC
coefficients ($\hat z_{ij}=\hat z_{ji}^*$)
\begin{equation}
\begin{array}{ccccccccc}
\hat z_{uc}&=& -s_5s_6e^{i\delta_3},&\hat z_{ut}&=&
-s_4s_6e^{i\delta_2},&
\hat z_{ct}&=& -s_4s_5e^{i(\delta_2-\delta_3)}\\
\hat z_{ux}&=& -s_6,& \hat z_{cx}&=& -s_5e^{-i\delta_3},&
\hat z_{tx}&=& -s_4e^{-i\delta_2}.
\end{array}
\end{equation}

\medskip\noindent\underline{(b) Unitarity constraints}

Unitarity constraints on the $3\times 3$ CKM matrix give linear
three-term relations that can be expressed graphically as triangle
relations in the complex plane; see Fig.~\ref{fig:quad}.
With $4\times 4$ mixing, they become four-term relations; e.g. for
one $Q=-{1\over 3}$ singlet, we have
\begin{equation}
V^*_{ui}V_{uj}+V^*_{ci}V_{cj}+V^*_{ti}V_{tj}+V^*_{oi}V_{oj}=
\delta_{ij},
\label{eq:quad1}
\end{equation}
or again,
\begin{equation}
V^*_{id}V_{jd}+V^*_{is}V_{js}+V^*_{ib}V_{jb}+V^*_{ix}V_{jx}=
\delta_{ij}.
\label{eq:quad2}
\end{equation}
For $i\ne j$ these are expressible as
quadrangle conditions in the complex plane.  The first three terms
in each case, however, are precisely the three sides of a triangle
if CKM unitarity holds (the most discussed example is
Eq.(\ref{eq:quad1}) with $i=b,j=d$). Thus $4\times 4$ unitarity replaces the
CKM triangle relations by quadrangle relations.  In
Eq.(\ref{eq:quad1}) the fourth side of the quadrangle is
$V^*_{oi}V_{oj}=-z_{ij}$, the FCNC coefficient\cite{nir}.  In
Eq.(\ref{eq:quad2}) the fourth side is $V^*_{ix}V_{jx}$, that
occurs in certain flavor-changing box diagrams (see below).

In the case of one $Q=-{1\over 3}$ singlet quark, the squares of the
elements in each row and column of the $4\times 4$ unitary matrix $V$
sum to unity.  Hence the experimental lower bounds on the CKM
submatrix elements\cite{gilman} shown in Eq.(\ref{eq:ckmlim}) give
constraints:
\begin{equation}
\begin{array}{ccccccccc}
|V_{ux}| &\alt& 0.08,&\quad |V_{cx}| &\alt& 0.57, &\quad |V_{tx}| &\alt& 1.0,\\
|V_{od}| &\alt& 0.15,&\quad |V_{os}| &\alt& 0.56, &\quad |V_{ob}| &\alt&
1.0.\end{array}
\end{equation}
Also each quadrangle must close, so the exotic fourth side is bounded by
the sum of the upper limits of the three conventional CKM sides, giving
\begin{equation}
\begin{array}{ccccccccc}
|V_{ux}||V_{cx}| &\alt& 0.44,&\quad |V_{ux}||V_{tx}| &\alt& 0.15,
&\quad |V_{cx}||V_{tx}| &\alt& 0.60,\\
|V_{od}||V_{os}| &\alt& 0.45,&\quad |V_{od}||V_{ob}| &\alt& 0.03,
&\quad |V_{os}||V_{ob}| &\alt&0.61 .
\end{array}
\end{equation}
Finally, when eventually we obtain upper bounds on $|V_{oj}|$
($j=d,s,b$) from other data, unitarity will imply a lower bound on
$|V_{ox}|^2 = 1-\Sigma_j|V_{oj}|^2$, and hence an upper bound on
$\Sigma_i|V_{ix}|^2 = 1-|V_{ox}|^2$ ; this latter bound will apply
equally to each $|V_{ix}|^2$ in the summation ($i=u,c,t$). See
Section VI(e) below.

In the case of one $Q={2\over 3}$ singlet quark, bounds on the CKM
submatrix elements of the mixing matrix $\hat V$ of
Eq.(\ref{eq:defvhat}) give analogous constraints:
\begin{equation}
\begin{array}{ccccccccc}
|\hat V_{uo}| & \alt & 0.08, &\quad |\hat V_{co}| & \alt & 0.57, &\quad
|\hat V_{to}| & \alt & 1.0, \\
|\hat V_{xd}| & \alt & 0.15, &\quad |\hat V_{xs}| & \alt & 0.56, &\quad
|\hat V_{xb}| & \alt & 1.0, \\
|\hat V_{uo}||\hat V_{co}| &\alt& 0.44,
&\quad |\hat V_{uo}||\hat V_{to}| &\alt& 0.15,
&\quad |\hat V_{co}||\hat V_{to}| &\alt& 0.60,\\
|\hat V_{xd}||\hat V_{xs}| &\alt& 0.45,
&\quad |\hat V_{xd}||\hat V_{xb}| &\alt& 0.03,
&\quad |\hat V_{xs}||\hat V_{xb}| &\alt& 0.61 .
\end{array}
\end{equation}

\section{Direct singlet-quark production constraints}

\noindent\underline{(a) $Z$ decays}

At $e^+e^-$ colliders, $\bar xx$ pairs can be produced directly via
their $\gamma$ and $Z$ couplings, and $x\bar q$ or $\bar xq$ pairs
via  FCNC. The most stringent bounds at present come from the observed
$Z$ decay widths,  from which it appears that contributions beyond
the SM are limited by~\cite{lep}
\begin{equation}
 \Gamma_Z(\hbox{non-SM}) \alt 15\rm\ MeV \;.
\end{equation}
For the case $Q_x = -{1\over3}$, the partial widths for decay to one
light plus one new quark are
\begin{equation}
\Gamma(Z\to\bar dx) = \Gamma(Z\to d\bar x) =
3\Gamma_Z^0 K \left|z_{dx}\right|^2 F_x =
(0.66\ {\rm GeV}) \left|z_{dx}\right|^2 F_x ,
\end{equation}
and for the case $Q_x={2\over3}$ we have
\begin{equation}
\Gamma(Z\to\bar ux) = \Gamma(Z\to u\bar x) =
3\Gamma_Z^0 K \left|\hat z_{ux}\right|^2 F_x =
(0.66\ {\rm GeV}) \left|\hat z_{ux}\right|^2  F_x ,
\end{equation}
where $\Gamma_Z^0=G_F M_Z^3 \Big/ \left(12\pi\sqrt 2\right)=0.17$~GeV ,
$F_x= \left(1-m_x^2/M_Z^2\right)^2 \left(1+m_x^2/ 2M_Z^2\right)$
and $K = 1 + (8\pi /9)\alpha_s(M_Z) = 1.33$ is a QCD factor.
For each  $x\bar q + \bar xq$ contribution to remain within the bound
on  $\Gamma_Z$(non-SM) sets $m_x$-dependent constraints on the FCNC
coefficients,
\begin{eqnarray}
\sqrt{F_x} \left|z_{ix}\right| &\alt& 0.11 \quad i=d,s,b \;, \quad
(Q_x=-\case1/3) \; \\
\sqrt{F_x}  \left|\hat z_{jx}\right| &\alt& 0.11 \quad j=u,c \;,
\quad
(Q_x=\case2/3)
\;.
\end{eqnarray}

The partial widths for decays to $\bar xx$ pairs are
\begin{equation}
\Gamma(Z\to \bar xx) = 24\,\Gamma_Z^0\, K\,\left[ 1-4m_x^2 /
M_Z^2  \right]^{1/2}
\left[ g_V^2(1+2m_x^2/M_Z^2) + g_A^2 (1-4m_x^2/M_Z^2)\right]
\end{equation}
where
$$\begin{array}{ccccccc}
g_V&=&-{1\over4 }z_{xx}+{1\over 3}\sin^2\theta_W,&\quad
g_A&=&-{1\over4 }z_{xx}& (Q_x = -{1\over 3}),    \\
g_V&=& {1\over4 }\hat z_{xx}-{2\over 3}\sin^2\theta_W,&\quad
g_A&=& {1\over4 }\hat z_{xx}& (Q_x = {2\over 3})\;.
\end{array}$$
In the limit of small singlet-doublet mixing, we have $z_{xx}\simeq0$
or $\hat z_{xx}\simeq0$ and hence $g_V\simeq -Q_x\sin^2\theta_W$,
$g_A\simeq 0$.  In this limit the upper bound $\Gamma(Z\to \bar xx) <
\Gamma_Z$(non-SM) gives
\begin{eqnarray}
m_x &\agt& 42\ {\rm GeV} \quad \left(Q_x=-\case1/3\right)
\;,
\label{eq:lepxd}\\
m_x &\agt& 45\ {\rm GeV} \quad \left(Q_x=\case2/3\right) \;.
\label{eq:lepxu}
\end{eqnarray}
Figure~\ref{fig:xx} shows the corresponding $\bar xx$ contribution
to $\Gamma_Z$(non-SM) versus $m_x$ for $Q_x=-\case1/3$ and
$\case2/3$. Direct searches at LEP for typical heavy quark signals
($t\to bW^{*+}$, $b'\to cW^{*-}$), based simply on event shapes,
set early limits $m_t > 44.5$ GeV and $m_{b'} > 45.2$ GeV
\cite{opal}, corresponding to upper limits  $\Gamma (Z\to
\bar b'b', \bar tt) < 20-30$ MeV.  Applying these limits to
singlet quarks gives a weaker result than Eq.(\ref{eq:lepxd})
and about the same as Eq.(\ref{eq:lepxu}); some improvements could
presumably be achieved with present much higher luminosities.

\medskip
\noindent\underline{(b) Hadroproduction}

At hadron colliders, $\bar xx$ pairs can be produced via QCD
interactions exactly like SM quark pairs. Their $x\to q'W$
CC decays into lighter quarks give signals rather
similar to the $t\to bW$ signals that have been looked for in
top-quark searches~\cite{cdf,dzero,cdf2}, although the details
may differ; they also have new decays into $qZ$ and/or $qH$.
We briefly discuss some examples.

\noindent
(i) For a heavy $Q=\case2/3$ singlet $x_t$ that mixes mostly with $t$
and has $M_W<m_{x_t}<M_W+m_t$, the dominant decay mode is $x_t\to bW$
while $x_t\to tZ,tH$ are kinematically forbidden. Hence the $\bar x_t
x_t$ signals look {\em exactly} like $\bar tt$ signals, including the
presence of taggable $b$-jets in the final state. Lower bounds on
$m_t$ such as the D0 result\cite{dzero}  $m_t > 131$ GeV
apply also to $m_{x_t}$.  Recently published evidence
for $t\bar t$ production\cite{cdf} could in principle be interpreted
as $\bar x_t x_t$ production, but electroweak radiative corrections
\cite{ewrc} already indicate a top mass near the observed value,
making $t\bar t$ production the most likely interpretation.
However, if there is an excess of top-type events above the SM
rate~\cite{cdf}, this could be due to $\bar x_t x_t$ production in
addition to $\bar tt$ production \cite{bp2}.

\noindent
(ii) A $Q=-\case1/3$ singlet $x_b$ that mixes mostly with $b$ and has
$M_Z<m_{x_b}<m_t+M_W$ would decay dominantly via $x_b\to bZ,bH$
with the $tW$ mode suppressed, escaping the usual top searches but
offering new $Z$ and  $H$ signals. If the latter are suppressed
(e.g., if $m_H>m_{x_b}$), early CDF limits on the remaining $Z$
signals imply a bound $m_{x_b}>85$~GeV~\cite{dp}.
This scenario gains fresh interest~\cite{bp2} from hints of possible
excess tagged $Z$ plus four jet events at the Tevatron~\cite{cdf}.

To be quantitative about signal expectations with $b$-tagging, let us
consider $x_b$ and $t$ to be degenerate ($m_{x_b}\simeq m_t$) for
simplicity, so that they are produced equally.  If $x_b$ is lighter
than this, the singlet signal rates will be correspondingly higher.
For the singlet decay we consider two extreme scenarios:
(A)~$m_H> m_{x_b}> M_Z$, so that $\Gamma(x_b\to bZ)\gg
\Gamma(x_b\to bH)$ and (B)~$m_H\simeq M_Z$ with small $b$-$x_b$
mixing, so that $\Gamma(x_b\to bZ)\simeq \Gamma(x_b\to bH)$.
For $b$-tagging efficiency, we assume $\epsilon_b=0.2$ to be the
probability for tagging a single $b$-quark; then the probability
for tagging a $b\bar b$ event is $1-(1-\epsilon_b)^2=0.36$ and the
probability for tagging a $b\bar bb\bar b$ event is
$1-(1-\epsilon_b)^4=0.59$. Our discussion is simplified in that we
neglect fake tags, assume $b$-tags are uncorrelated, and assume 100\%
acceptance. Then the probabilites for different final state
configurations including $b$-tagging are
$$\begin{array}{llc}
           & channel                          & probability\; with\;
tag \\
\bar bbWW \to&\bar bb(\ell\nu)(jj)    & 0.29\times 0.36 =
0.104 \\
\bar bbZZ \to&\bar bb(\ell\ell)(jj,bb)& 0.094\times 0.41
=0.039 \\
\bar bbZH \to&\bar bb(\ell\ell)(bb)   & 0.067\times 0.59
=0.040
\end{array}$$
summing over $\ell = e,\mu$ channels.   The first numerical factor
on the right is the branching fraction and the second factor is
the $b$-tagging probability. Thus the leptonic $W/Z$ event ratios
in our two $m_{x_b}\simeq m_t$ scenarios (A) and (B) are
\begin{equation}
N(tt\to W_{\ell\nu}+4j\, {\rm with\; tag})/N(xx\to Z_{\ell\ell}+4j\,
{\rm with\;
tag}) \simeq 2.7 (A)\quad or \quad 3.5 (B) .
\end{equation}
In contrast, the QCD electroweak background ratio is~\cite{bmps}
\begin{equation}
N(QCD\to W_{\ell\nu}+4j\, {\rm with\; tag})/N(QCD\to Z_{\ell\ell}+4j\,
{\rm with\; tag}) \simeq 10-14.
\end{equation}

\noindent
(iii) A $Q=-\case1/3$ singlet quark $x_d$ mixing mostly with $d$
would decay by $x\to uW,dZ,dH$ in the ratios $2:1:1$ modulo phase
space factors.  Thus for $m_x \gg M_W,M_Z,M_H$ the top-like signals
would be reduced roughly by a factor 2 for single-lepton channels
and by a factor 4 for dilepton signals, compared to a top quark of
the same mass; however, for smaller $m_x$ the reduction
is generally less, and in the window $M_W < m_x < M_Z, M_H$ there
is no reduction. But there is now no $b$-quark to tag. Examination
of earlier top-quark searches without a $b$-tag~\cite{dzero,cdf2},
scaling down the top-quark expectations by some factor between
1 and 4, shows that the range $M_W < m_x < M_Z, M_H$ is definitely
excluded, and probably some adjacent ranges of $m_x$ too, but more
cannot be said without detailed analysis.
Similar conclusions apply to $x_s$  singlets mixing mostly with $s$
and to charge $\case2/3$ singlets $x_u$ or  $x_c$ mixing
mostly with $u$ or $c$; there are small differences between these
cases, such as the lepton spectrum~\cite{bp} and the taggability
of $c$-quarks,  but they do not change the overall conclusion.

\noindent
(iv) Decays outside the detector.  The distinctive $x$-quark
signals will be lost if $x$ decays outside the detector.  If we
assume typical Lorentz factors $\beta\gamma\sim 2$ and require
that the mean decay distance $\ell_D = \beta\gamma c/\Gamma$
due to any single $x\to q_iW,q_jZ$ decay mode exceeds one
metre, Eqs.(\ref{eq:gccxd})-(\ref{eq:gfcncxd}) give
\begin{equation}
|V_{ix}| \alt 1.2\times 10^{-8} \left [{200\;
                   {\rm GeV}\over m_x}\right ]^{3\over 2},\quad
|V_{oj}| \alt 0.9\times 10^{-8} \left [{200\;
                   {\rm GeV}\over m_x}\right ]^{3\over 2} \; ,
\end{equation}
for $Q=-{1\over 3}$ , and similarly
\begin{equation}
|\hat V_{xi}| \alt 1.2\times 10^{-8} \left [{200\;
                        {\rm GeV}\over m_x}\right ]^{3\over 2},\quad
|\hat V_{oj}| \alt 0.9\times 10^{-8} \left [{200\;
                        {\rm GeV}\over m_x}\right ]^{3\over 2} \; ,
\end{equation}
for $Q={2\over 3}$.   If these conditions hold for all light
quark flavors $i,j$ , then singlet decay signals at hadron
colliders will be greatly suppressed.  [The conditions are
somewhat weaker for $m_x \alt M_W, M_Z$].

\medskip\noindent
\underline{(c) Leptoproduction}

At $ep$ colliders, singlet quarks can be produced by the same
$\gamma g$ fusion processes as SM quarks; $Zg$ fusion is also
possible (at reduced rates due to reduced $Z\bar xx$ couplings)
but $Wg$ fusion is only possible via mixing.  However, the reach
of the HERA collider for new quark detection is much less
that that of the Tevatron~\cite{hera} so this is not a promising
avenue for singlet discovery.

\medskip\noindent
\underline{(d) Summary}

The LEP mass bounds Eqs.(\ref{eq:lepxd})-(\ref{eq:lepxu}) are
virtually unconditional.  Hadroproduction bounds are much
stronger in particular cases [e.g. $m_x > 85$ GeV ($131$ GeV)
if the decays $x\to qZ$ ($x\to qW$) dominate completely], but
assume implicitly that the mixing with lighter quarks is not so
extremely weak that $x$ decays outside the detector [typified by
off-diagonal 4th row and column mixing-matrix elements all being
$\alt\; 10^{-8}[(200\;GeV)/m_x)^{3\over 2}$].

\section{Neutral meson-antimeson oscillations}

The existence of new singlet quarks can affect neutral
meson-antimeson oscillations in two different ways, through FCNC
tree-level $Z$ exchange and through box diagrams,
illustrated in Fig.~\ref{fig:treebox} for the $B_d^0$-$\bar
B_d^0$ case. In the tree diagram of Fig.~\ref{fig:treebox}(a), the
effects are due to an additional $Q=-\case1/3$ singlet that
generates the FCNC couplings of
Eqs.~(\ref{eq:Lfcnc})--(\ref{eq:dzij}); more generally, such FCNC
effects of $Q=-\case1/3$ singlets occur also for $K^0$-$\bar K^0$
and $B^0_s$-$\bar B^0_s$ oscillations, whereas analogous effects
from $Q=\case2/3$ singlets give $D^0$-$\bar D^0$ oscillations.

In the box diagram of Fig.~\ref{fig:treebox}(b), the effects come from
an additional $Q=\case2/3$ heavy quark option in the loop, along with
corresponding reductions in the original three generation couplings.
Similar effects are present in $K^0$-$\bar K^0$ and $B^0_s$-$\bar B^0_s$
oscillations, while analogous effects from $Q=-\case1/3$ singlets
occcur in the $D^0$-$\bar D^0$ case. The association of singlets with
their mixing effects is summarized in Table~\ref{mixing}.

\medskip\noindent\underline{(a) $Z$-exchange contributions}

We first analyze the $Z$-exchange FCNC effects, which are potentially
the most interesting. For $B^0_d$-$\bar B^0_d$ oscillations, the
contribution is
\begin{equation}
|\delta m|  = {\sqrt2 G_F m_B f_B^2 B_B \eta _B\over 3} |z_{db}^2|\;,
\label{Zmed}
\end{equation}
where $f_B$ is the $B_d$ decay constant, $B_B$ is the bag factor
($B=1$ is the vacuum saturation approximation) and
$\eta _B\approx 0.55$ is a QCD factor. (We assume the QCD correction
is the same for both the $Z$-exhange contributions and for the box
diagram contributions described below.) The analogous expressions for
$K^0$-$\bar K^0$, $D^0$-$\bar D^0$, $B^0_s$-$\bar B^0_s$ oscillations
involve $Re(z_{ds}^2),\ \hat z_{uc}^2,\ z_{sb}^2$, respectively. Actually
this FCNC process contributes coherently with the SM box diagrams. We
shall here assume very conservatively that the singlet-quark
$Z$-exchange contributions do not exceed the measured values. Then
from the measurements~\cite{pdg}
$|\delta m|_K    =(3.51\pm 0.02)\times 10^{-12}\,$MeV,
$|\delta m|_D    < 1.3          \times 10^{-10}\,$MeV,
$|\delta m|_{B_d}=(3.4 \pm 0.4 )\times 10^{-10}\,$MeV   we obtain
\begin{equation}
|Re(z_{ds}^2)| = |(Re\, z_{ds})^2-(Im\, z_{ds})^2| \alt 9\times10^{-8} \;,
\label{eq:zsd1}
\end{equation}
\begin{equation}
|\hat z_{uc}| = |\hat V_{uo}|\;| \hat V_{co}| \alt 9\times10^{-4} \;,
\label{eq:zcu1}
\end{equation}
\begin{equation}
|z_{db}| = |V_{od}|\;| V_{ob} | \alt 8\times10^{-4} \;,
\label{eq:zbd1}
\end{equation}
where the $\alt$ symbol reflects some uncertainties in the factors
$f,\ B,\ \eta$. Similar bounds on $z_{ds}$ and $z_{db}$ are given in
Ref.\cite{branco1,nir,cs}) and on $\hat z_{uc}$ in Ref.\cite{branco4}.
For $|\delta m|_{B_s}$ there is only an experimental upper
limit~\cite{pdg} and hence no bound on $|z_{sb}|=|V_{os}|\,|
V_{ob}|$. We take the $D$ and $B$ decay constants from
Narison~\cite{narison}:  $f_D=1.37f_\pi, f_B=1.49f_\pi$, with
$f_\pi=0.131$~GeV and $f_K=0.160$~GeV, and set  $B=1$ and $\eta=0.55$ in
all cases.  Taking the lower bound $B={1\over 3}$
instead would raise the limits above by a factor $\sqrt 3$.

\medskip\noindent\underline{(b) New box diagram contributions}

In the case of box diagram contributions, the constraints on the
mixing are rather different. First consider the case of
$B^0_d$-$\bar B^0_d$ box diagrams with an additional $Q_x=\case2/3$
contribution, with $m_x\simeq m_t$ approximate degeneracy. Then
the SM formula is
\begin{equation}
|\delta m|_{\rm SM}  = {G_F^2 Bf_B^2 m_B \eta_B \over 6\pi^2}
\left|V_{td}V^*_{tb}\right|^2_{\rm CKM} \left|I_B\right| \;, \label{SMform}
\end{equation}
where $I_B$ is a box-integral factor (see e.g.\ Ref.~\cite{book}),
and the effect of adding an extra singlet is to replace the CKM
factor by
\begin{equation}
\left| \hat V_{xd} \hat V_{xb}^* + \hat V_{td} \hat V_{tb}^*
\right|^2 =
\left| V_{ud} V_{ub}^* + V_{cd} V_{cb}^*\right|^2_{\rm CKM}
\end{equation}
using unitarity. However, $|V_{td} V_{tb}^*|_{CKM} = |V_{ud} V_{ub}^*
+ V_{cd} V_{cb}^*|_{\rm CKM}$, so the prediction for $|\delta m|$ is
effectively unchanged in this $x,t$ mass-degenerate limit. Only if
$x$ is much heavier than $t$ can significant changes arise. Similar
conclusions apply to $K^0$-$\bar K^0$ oscillations.

The $Q=-\case1/3$ box diagram contributions to $D^0$-$\bar D^0$
mixing are potentially more interesting, because
$d,\ s$ and $b$ are relatively light compared to the allowed mass
scale for $x$. Here the $x$ contributions may be dominant
(depending on the size of the mixing) and given by
\begin{equation}
|\delta m|_D = {G_F^2 B f_D^2 m_D \eta_D\over 6\pi^2} \left|
               V_{cx} V_{ux}^* \right|^2 \left|I_D\right| \;,
\end{equation}
where $|I_D|\simeq m_x^2$ for $m_x\sim200$~GeV, giving
\begin{equation}
| V_{cx}|\;| V_{ux}| \alt 0.7\times10^{-2} \left(200{\rm\
GeV}\over m_x\right) \;.
\end{equation}
A similar bound is noted in Ref.~\cite{babu} for mixing a
fourth-generation $b'$ quark, that is essentially equivalent to
singlet mixing in this context.  It is expected that a future sample
of $10^8$ reconstructed $D$'s  would have a factor 20 improvement in
sensitivity to $\delta m_D$, and would consequently give a factor
$\sim 4-5$ more sensitivity to the above mixing. Note that SM short-
and long-distance contributions are far below this
sensitivity~\cite{pak,hew}.

The parameter $\epsilon_K$ , that describes CP-violation in $K^0-\bar
K^0$ oscillations, also receives tree-level $Z$-exchange
contributions from $Q=-{1\over 3}$ singlet mixing:
\begin{equation}
|\epsilon_K|  = {G_F m_K B_K f_K^2 \over 12|\delta m_K|}
|Im(z_{ds})^2|.
\end{equation}
Requiring $|\epsilon_K| \le |\epsilon_K|_{exp} =
2.27\times 10^{-3}$ gives the bound\cite{nir,cs}
\begin{equation}
|Im(z_{ds})^2| \alt 6\times 10^{-10},\quad
|Re(z_{ds})\, Im(z_{ds})| \alt 3\times 10^{-10} .
\label{eq:imzds}
\end{equation}
Combined with Eq.(\ref{eq:zsd1}), this gives
\begin{equation}
|z_{ds}| = |V_{od}|\, |V_{os}| \alt 3\times 10^{-4} .
\label{eq:zsdosc}
\end{equation}

\section{FCNC decays and FDNC effects}

\medskip\noindent\underline{(a) $K_L\to\mu^+\mu^-$ decay}

Experimental measurements on FCNC decays imply constraints on the
FCNC $Z$ couplings and hence on singlet-quark mixing parameters
\cite{bdpw,rizzo,branco1,nir,branco2}.  For example,
$K_L\to\mu^+\mu^-$ has a $Z$-mediated diagram if a $Q=-{1\over 3}$
singlet $x$ mixes with $d$ and $s$, contributing the decay width
\begin{equation}
\Gamma (K_L\to \mu^+\mu^-)_Z = {2G_F^2f_K^2m_Km_{\mu}^2\over 8\pi}
  [1-4m_{\mu}^2/m_K^2]^{1/2}
  [({1\over 2}-\sin^2\theta_W)^2 + (\sin^2\theta_W)^2]\; |z_{ds}|^2.
\label{eq:klmumu}
\end{equation}
After subtracting the contribution for the $\gamma\gamma$ intermediate
state (an imaginary decay amplitude), the latest Brookhaven results
\cite{e791} indicate an upper limit $B_{real} < 5.6\times 10^{-10}$
($90\%$ CL) on the contribution to the branching fraction from the
real part of the decay amplitude. Using this to bound the contribution
from $Re(z_{ds})$ we obtain
\begin{equation}
|Re(z_{ds})| \alt 0.64 \times 10^{-5}.
\end{equation}
The combined bound on $|z_{ds}|$ remains unchanged.

\medskip\noindent\underline{(b) $B^0,D^0\to\mu^+\mu^-$ decays}

 Analogous formulas describe the tree-level
contributions to $D^0\to\mu^+\mu^-$ and $B^0\to\mu^+\mu^-$ decays
(without the factor 2 because $D^0$ and $B^0$ are not pure $CP=-1$
states).  Requiring that the $Z$-exchange contributions are within
the experimental limits
$B(D^0 \to \mu^+ \mu^-) < 1.1 \times 10^{-5}$ and
$B(B^0 \to \mu^+ \mu^-) < 5.9 \times 10^{-6}$ \cite{pdg} ,
gives the constraints
\begin{equation}
|\hat z_{uc}| = |\hat V_{uo}|\;|\hat V_{co}| \alt 0.20 \;,
\label{eq:zcu2}
\end{equation}
\begin{equation}
|z_{db}|     =  | V_{od}    |\;|     V_{ob}| \alt 0.04 \;,
\label{zbd2}
\end{equation}
much weaker than the oscillation bounds
Eqs.(\ref{eq:zcu1})-(\ref{eq:zbd1}).

\medskip\noindent\underline{(c) $B,D\to X\ell^+\ell^-$ decays}

The rare decays $B\to X\ell ^+\ell ^-$  occur at tree level, via
FCNC couplings $z_{db}$ and $z_{sb}$ which give
\begin{equation}
{\Gamma(B\to\ell^+\ell^-X) \over \Gamma(B\to\ell^+\nu X)} =
[({1\over 2}-\sin^2\theta_W)^2 + \sin^4\theta_W]\times
{|z_{db}|^2 + |z_{sb}|^2 \over |V_{ub}|^2 + \rho |V_{cb}|^2},
\label{eq:bllx}
\end{equation}
where $\rho \simeq 0.5$ is a phase space factor;
$\rho = 1 - 8r^2 + 8r^6 - r^8 - 24r^4\ln(r)$ with
$r= m_c/m_b=0.316\pm 0.013$.   Hence the experimental limit
$B(B\to X\mu ^+\mu ^-)\le 5.0\times 10^{-5}$ gives the
constraints\cite{nir,branco2}
\begin{equation}
|z_{db}| = |V_{ob}|\; |V_{od}|\;\alt\; 0.04\times |V_{cb}|\;
\alt\; 2\times 10^{-3}\; ,
\label{eq:zbd3}
\end{equation}
\begin{equation}
|z_{sb}| = |V_{ob}|\; |V_{os}|\;\alt\; 0.04\times |V_{cb}|\;
\alt\; 2\times 10^{-3}\; .
\label{eq:zbs}
\end{equation}
The first bound is competitive with that from $B_d-\bar B_d$
oscillations in Eq.(\ref{eq:zbd1}).  Upper limits have recently been
given for some $D\to\mu^+\mu^-+{\rm hadrons}$ branching fractions
\cite{e653}, suggesting an inclusive upper limit of order $(1-2)\times
10^{-3}$ (although no explicit value is quoted); such a limit would
however only give $|\hat z_{uc}|\alt 0.2-0.3$, possibly competitive
with Eq.~(\ref{eq:zcu2}) but much weaker than Eq.~(\ref{eq:zcu1}).

\medskip\noindent\underline{(d) $B\to s(d)\gamma$ decays}

The rare decays $B\to s(d)\gamma $ have also been
considered\cite{branco3,handoko}.
In the SM they go via $W$-loop diagrams; adding a down-type singlet
quark introduces new $Z$-loop diagrams, using the FCNC couplings
$z_{ij}$ ($H$-loops are usually negligible).   These can be incorporated
into the conventional analyses by adding their contributions
into the coefficients  of the effective operators
of the magnetic and chromomagnetic moment couplings
$f_{\gamma}^{(1)}$ and $f_g^{(1)}$
as described in Appendix A.   The ratio of
$\Gamma (b \rightarrow q \gamma)$ (where $q=d,\; s$)
to the inclusive semileptonic decay width is then given by
\begin{equation}
{{\Gamma(b\rightarrow q \gamma)}\over {\Gamma (b\rightarrow ce\nu )}}
={{6\alpha }\over {\pi \rho \lambda }}
{{|V_{tq}^*V_{tb}|^2}\over {|V_{cb}|^2}}|c_7(m_b)|^2 \;,
\label{ratio}
\end{equation}
where $\alpha $ is the electromagnetic coupling and
\begin{eqnarray}
c_7(m_b)=\left [{{\alpha _s(M_W)}\over {\alpha _s(m_b)}}\right
]^{16/23}
\Bigg \{c_7(M_W) &-& {8\over 3} c_8(M_W)\left [1-\left (
{{\alpha _s(m_b)}\over {\alpha _s(M_W)}}\right )^{2/23}\right ]\Bigg
\}
\nonumber \\
&+&\sum _{i=1}^8 h_i\left ({{\alpha _s(M_W)}\over {\alpha _s(m_b)}}
\right )^{a_i}
\;.
\end{eqnarray}
The Wilson coefficients $c_7$ and $c_8$, the coefficients $h_i$, and
the exponents $a_i$
from the $8\times 8$ anomalous dimension matrix\cite{qcd} are given
in Appendix A. The phase-space factor $\rho $ is defined below
Eq.(\ref{eq:bllx}) and the QCD correction factor $\lambda $
for the semileptonic process is
$\lambda = 1 - {2\over 3} f(r,0,0)\alpha_s^{}(m_b)/\pi $
with $f(r,0,0)=2.41$\cite{QCDcorr}.
We remark that the FCNC diagrams include not only $Z$-loops
but also tree-level $Z$-exchanges between the b-quark and
the spectator antiquark in a decaying B-meson, not commented
upon in previous literature.  However, these $Z$-exchanges are
suppressed relative to $Z$-loops by factors $f_B/m_B \sim 1/25$
in decay amplitudes\cite{fbmb}, so we do not pursue them here.

In the SM one expects the ratio
$B(b\to d\gamma )/B(b\to s\gamma)\approx |V_{td}/V_{ts}|^2$, since the
QCD corrections largely cancel out.  The additional FCNC terms are
proportional to $z_{qb}/(V_{tb}V^*_{tq})$  in each case ($q=d,s$),
and it has been shown that\cite{branco3}
\begin{eqnarray}
\left |{{z_{db}}\over {V_{tb}V_{td}^*}}\right |\le 0.93\;, \label{zbd093}\quad
\left |{{z_{sb}}\over {V_{tb}V_{ts}^*}}\right |\le 0.04\;.
\end{eqnarray}
These limits permit singlet quarks to have greater impact on the
$b\to d\gamma$ rate (e.g. if $z_{db}\sim z_{sb}$).  On the other hand,
one expects from the general decoupling theorem\cite{bowick} that
$z_{db}$ is much smaller than $z_{sb}$.

An up-type singlet quark can also be considered. Its contribution
is the same as from a standard fourth generation, giving
\begin{equation}
{{\Gamma(b\rightarrow q \gamma)}\over {\Gamma (b\rightarrow ce\nu )}}
={{6\alpha }\over {\pi \rho \lambda }}
\left ({{|\hat V_{tq}^*\hat V^{}_{tb}c_7^t(m_b)
+         \hat V_{xq}^*\hat V^{}_{xb}c_7^x(m_b)|^2}
          \over {|V_{cb}|^2}}\right )\;,
\label{eq:bsgup}
\end{equation}
where the matching conditions for the relevant Wilson coefficients
are again given in Appendix A.  The major contributions to
$B(b\to q\gamma)$ are now the $t$- and $x$-quark loop terms in
Eq.(\ref{eq:bsgup}). Notice that $\Gamma(b\to q\gamma)$ is the same
as in the SM when  $m_x=m_t$, for the same reason as
in $B^0_d$-$\bar B^0_d$ and $K^0$-$\bar K^0$ oscillations above.
But if $m_x$ deviates significantly from $m_t$, an enhancement or
suppression relative to the SM can be expected (as with a fourth
generation \cite{hss}).

Figure~\ref{fig:bsgam} shows the
$b\to s\gamma$ rate versus $m_x$ with various values of
$|\hat V_{xs}^*\hat V^{}_{xb}|$,  for the SM plus one up-type singlet
quark.  We have assumed here that the phase of
$\hat V_{xs}^*\hat V^{}_{xb}$ is the same as that of
$\hat V_{ts}^*\hat V^{}_{tb}$ within a sign,  so that deviations from
the SM are maximized.   We note incidentally that the unitarity
constraint on $|\hat V_{xd}||\hat V_{xb}|$ helps to guarantee that
$B(b\to d\gamma)$ with a $u$-type singlet remains close to the SM.

\medskip\noindent\underline{(e) $Z$ decays}

We turn now to FDNC effects.  At tree level, introducing mixing with
a singlet quark $x$ simply reduces the left-handed coupling of a
conventional quark $i$ by a factor
$1 - |V_{oi}|^2 /(1-{2\over 3}\sin^2\theta_W)$
(for charge $Q_x=Q_i=-{1\over 3}$), or by a factor
$1-|\hat V_{io}|^2 /(1-{4\over 3}\sin^2\theta_W)$
(for charge $Q_x=Q_i={2\over 3}$),  leaving right-handed
couplings unchanged;  see Eqs.~(\ref{eq:dfdnc}),(\ref{eq:ufdnc}).
We shall neglect singlet-mixing effects at one-loop level, where
they are small corrections to small corrections.

The $Z$ partial decay widths, branching fractions and asymmetry
measurements directly probe the FDNC $Zqq$ couplings.
$Z\to b\bar b$ decay is an interesting case to consider, since there
is at present some discrepancy between the LEP data \cite{lep}
and the SM prediction for the ratio
$R_b = \Gamma (Z\to b\bar b) /\Gamma(Z\to {\rm hadrons})$:
\begin{eqnarray}
R_b(\rm LEP) &=& 0.2202 \pm 0.0020\;, \qquad
R_b(\rm SM) = 0.2156 \pm 0.0004 \;.
\end{eqnarray}
Since $b-x$ mixing reduces the $Z\to b\overline{b}$ coupling, it
would make the discrepancy worse.  The decay width has the form
\begin{eqnarray}
\Gamma(Z\to \bar bb)=
{{\sqrt{2}G_FM_Z^3}\over {\pi}}\beta\left (\beta^2(g^b_A)^2
+{{3-\beta^2}\over 2}(g^b_V)^2\right ),
\end{eqnarray}
where $\beta$ is the CM velocity and
$g^b_A$ and $g^b_V$ are the axial and vector $Zbb$ couplings,
so down-type singlet mixing dilutes the tree-level contribution by a
factor $\approx (1-2.4|V_{ob}|^2)$.  It is inadvisable to derive a
limit on $V_{ob}$ from this result alone, however, since the SM
itself is on the verge of being excluded. Many models with
down-type singlets also give corrections to $Z\to b\bar b$
from mixing $Z$ with a new $Z^\prime$; these too are typically
negative\cite{nr}.

A global comparison of all FDNC effects with the latest LEP and SLC
data
leads to the following constraints (see final paper of
Ref.\cite{nr}):
$$
\begin{array}{cccccc}
|V_{od}|^2      < 0.0023\;,\quad &|V_{os}|^2      < 0.0036\;,\quad &
|V_{ob}|^2      < 0.0020\;,\quad  \\
|\hat V_{uo}|^2 < 0.0024\;,\quad &|\hat V_{co}|^2 < 0.0042\;,\quad &
\end{array}
$$
assuming at most one singlet quark mixes with each conventional
quark.  From these numbers, unitarity of $V$ then gives
\begin{equation}
|V_{ox}|\; > 0.996, \ |V_{qx}|\; < 0.089, \ (q=u,c,t).
\end{equation}

\medskip\noindent\underline{(f) Other FDNC effects}

Singlet mixing could also change FDNC effects in neutrino scattering
and atomic parity-violation measurements\cite{ll}, but there appear
to be no useful constraints from this quarter.

\section{CP asymmetries}

The amount of CP violation in the SM is measured  by the size of
the unitarity triangle in Fig.~\ref{fig:quad}. How this CP violation
shows up in decays is determined by the angles of the unitarity
triangle(s), which appear as CP asymmetries in decays to CP eigenstates.
The angles
\begin{eqnarray}
\beta \equiv {\rm arg}\left (-{{V_{cd}V_{cb}^*}\over {V_{td}V_{tb}^*}}\right )
\;,
\qquad
\alpha \equiv {\rm arg}\left (-{{V_{td}V_{tb}^*}\over {V_{ud}V_{ub}^*}}\right )
\;,
\end{eqnarray}
that characterize CP violation, are directly measurable in $B_d$
decays with $b\to c$ and $b\to u$ respectively.
The prototype processes for measuring $\beta $ and
$\alpha $ are $B_d\to \psi K_S$ and $B_d\to \pi ^+\pi ^-$ respectively.
[The angle  $\gamma \equiv {\rm arg}(-V_{ud}V_{ub}^*/V_{cd}V_{cb}^*)$ can
be measured in the decay $B_s\to \rho K_S$, which will prove much harder
at a $B$ factory because of the small branching fraction and the possible
contamination from penguin contributions.]
Present information on the third generation couplings does not
tell us much about the asymmetries.  Future improved measurements of the
CKM mixing angles  will pin down the SM prediction more precisely.
We find the biggest uncertainty in the SM asymmetries stems from the
uncertainty in $V_{ub}$, a quantity ripe for better measurement at a
$B$-factory.

We assume as usual that the asymmetries are dominated by the
interference between two amplitudes, one of which is given by
$B^0_d$-$\bar B^0_d$ oscillations with $\Gamma _{12}<<M_{12}$.
The time-dependent CP asymmetry in the decay of a $B^0_d$ or
$\overline{B}^0_d$ into some final CP eigenstate $f$ is
\begin{eqnarray}
{\Gamma(B^0_d(t)\to f)-\Gamma(\overline{B}^0_d(t)\to f)
\over \Gamma(B^0_d(t)\to f) + \Gamma(\overline{B}^0_d(t)\to f)}
=-{\rm Im}\; \lambda (B_d\to f)\,\sin (\delta m\;t)\;, \label{asymm}
\end{eqnarray}
where $\delta m$ is the (positive) difference in meson masses,
the mesons states evolve from flavor eigenstates $B^0_d$ and
$\overline{B}^0_d$ at a time $t=0$, and ${\rm Im}\; \lambda(B_d\to f)$ is the
time-independent asymmetry. The quantity
${\rm Im}\; \lambda(B_d\to f)$ is $-\sin 2\beta$
and $\sin 2\alpha $ for $f=\psi K_S$ and $f=\pi ^+\pi ^-$ respectively
in the SM (we neglect possible penguin diagrams in the decay
$B^0_d\to \pi ^+\pi ^-$.

We consider the allowed range for the Wolfenstein parameterization
involving $\rho $ and $\eta $ recently given in Ref.~\cite{al}.
The angles $\alpha $ and $\beta $ are easily related to $\rho $ and
$\eta $ through the unitarity triangle in Fig.~\ref{fig:quad}
\begin{eqnarray}
\sin 2\alpha&=&{{2\eta \left (\eta ^2+\rho (\rho -1)\right )}\over
{\left (\eta ^2+(1-\rho )^2\right )(\eta ^2+\rho ^2)}}\;, \\
\sin 2\beta&=&{{2\eta (1-\rho )}\over {\eta ^2+(1-\rho )^2}}\;.
\end{eqnarray}

In the presence of $d$-type singlet quarks the unitarity triangle becomes
a quadrangle as described in Section III, and the CP asymmetries in $B$
decays are altered from SM expectations. The deviations occur
in two ways.\\
(1) The angles $\beta $ and $\alpha $
no longer have SM values,
because the revised unitarity constraint yields different allowed ranges
and more general phases for the CKM elements.\\
(2) There is an additional $B_d-\bar B_d$ oscillation contribution from
tree-level $Z$-mediated graphs.\\
The asymmetry expressions are modified to
\begin{eqnarray}
{\rm Im}\; \lambda(B_d\to \psi K_S)&=&
-\sin \left (2\beta+{\rm arg}\Delta _{bd}\right )\;, \\
{\rm Im}\; \lambda(B_d\to \pi ^+\pi ^-)&=&
\sin \left (2\alpha+{\rm arg}\Delta _{bd}\right )\;,
\end{eqnarray}
where\cite{branco3}
\begin{eqnarray}
\Delta _{bd}&=&1+r_de^{2i\theta _{bd}}\;,\\
r_d&=&{{4\pi M_W^2\sin ^2\theta _W}\over {\alpha I_B(x_t)}}
\left |{{z_{bd}}\over {V_{td}V_{tb}^*}}\right |^2\;\simeq 140
\left |{{z_{bd}}\over {V_{td}V_{tb}^*}}\right |^2\;, \\
\theta _{bd}&=&{\rm arg}\left [{{z_{bd}}\over {V_{td}V_{tb}^*}}\right ]\;,
\end{eqnarray}
and $I_B(x_t=m_t^2/M_W^2)$ is the box integral(see e.g.\ Ref.~\cite{book})
\begin{eqnarray}
I_B(x_t)={1\over 4}M_W^2\left [x_t\left (1+{9\over {1-x_t}}
-{6\over {(1-x_t)^2}}
\right )-{{6x_t^3}\over {(1-x_t)^3}}\ln x_t\right ]\;.
\end{eqnarray}

The contribution of $z_{db}$ to the unitarity quadrangle can be described
by a magnitude and a phase $\theta_{bd}$ (relative to $V_{td}V_{tb}^*$).
This phase can take any value between $0$ and $2\pi $, but the magnitude
must be consistent with closure of the quadrangle. In Fig.~\ref{fig:beta}
we show the asymmetry for the decay $B_d\to \psi K_S$ for different
values of the parameters\cite{branco3}
\begin{eqnarray}
\delta _d\equiv \left |{{z_{bd}}\over {V_{td}V_{tb}^*}}\right |\;, \qquad
\theta _{bd}\;,\label{eq:defin}
\end{eqnarray}
with the CKM angles, the top mass and the $B$ lifetime
fixed at their central values:
$|V_{ud}|=0.9743$, $|V_{cd}|=0.204$, $|V_{ub}|=0.0035$,
$|V_{cb}|=0.40$,
$m_t=174$ GeV and $\tau _B^{}=1.50$ ps.
We take $\sqrt {Bf_B^2}=195$ MeV as we did previously, and use the
next-to-leading order value for the QCD correction $\eta_B=0.55$\cite{buras}.
By taking the coherent sum of the contributions to $B^0_d$-$\bar B^0_d$
mixing from Eqs.~(\ref{Zmed}) and (\ref{SMform}), and the the mixing parameter
$x_d=\delta m/\Gamma=0.71$, one can determine $|V_{td}V_{tb}^*|$.
The shaded band in Fig.~\ref{fig:beta} indicates the allowed range in
the SM for the asymmetry ${\rm Im}\; \lambda (B_d\to \psi K_S)$.

The quantity $\delta_d$ can be quite large, as indicated by
Eq.~(\ref{zbd093}), but Fig.~\ref{fig:beta} shows big effects
even with much smaller $\delta_d$.  One notices that the CP
asymmetry ${\rm Im}\; \lambda(B_d\to \psi K_S)$ is negative in the SM,
but with sufficiently large $\delta_d$ one can obtain
positive values\cite{nir,branco3}. The effect of singlet quarks on
CP asymmetries can therefore be dramatic\cite{nir,branco3}.

The CP asymmetry ${\rm Im}\; \lambda(B_d\to \pi ^+\pi ^-)$ is shown in
Fig.~\ref{fig:alpha} for various values of $\delta_d$ and
$\theta _{bd}$. Here the SM expectation covers the entire range,
so merely measuring the sign of the CP asymmetry could not upset the SM.
But given well-determined CKM elements, deviations from SM predictions
could be significant and could provide evidence for singlet quarks.

\section{GUT sources of singlet quarks}

\medskip\noindent\underline{(a) Generalities}

GUT models provide arguments for the existence of particles with
exotic quantum numbers, but also impose restrictions upon them.
In this Section, we explore the constraints on singlet-quark models
implied by coupling-constant unification and perturbativity. Most of
the examples we consider are supersymmetric models, and one must bear
in mind that these models have extra contributions to the processes
described above, so that the constraints obtained can be affected.

Singlet quarks considered alone do not introduce gauge (or gravitational)
anomalies, but they spoil the successful gauge coupling unification
of the minimal supersymmetric model (MSSM) if the singlets are below
the GUT scale, since they change the running of the $SU(3)$ and $U(1)$
couplings but not the $SU(2)$ coupling.  For down-type singlets, this can
be remedied by adding more fermions to fill out the ${\bf 5}$ and
${\bf 5^*}$ representations of $SU(5)$  or the ${\bf 10}$ of $SO(10)$;
see the examples below.

For up-type singlets, however, it is harder to find a consistent
scenario, if one believes that gauge coupling unification is due to a
GUT symmetry and therefore wants to retain the desert between the
GUT scale and the scale of the exotic fermions.  The model of Barbieri
and Hall\cite{bh} postulates that singlet quarks arise as supersymmetric
partners (gauginos) of gauge bosons from a unification group that assigns
a fourth color to leptons, so here the singlet with the right quantum
numbers to mix with the top quark is not a matter fermion at all.
We can introduce top-like singlets as matter fermions
by assigning them to the adjoint representation of the GUT group.  The
smallest suitable representation of matter fermions is then
the ${\bf 45}$ of $SO(10)$, or the ${\bf 78}$ of $E_6$.
But these representations are too large; they destroy the asymptotic
freedom of the strong coupling, and contain extra doublet quarks besides.
Alternatively, in the context of $SU(5)$, we can introduce one
up-type singlet quark by adding one extra light $\bf 10$ and one
$\bf 10^*$ representation; these bring one extra vector-singlet lepton
plus a vector-doublet of quarks too, and restore gauge unification with
$b_3=0$ at one-loop level.  Two-loop effects become large, however,
and large threshold corrections must be invoked to restore
gauge coupling unification.  Apart from this ${\bf 10+10^*}$ scenario,
there appears to be no simple way to arrive at a low-energy model with
up-type singlet quarks from a desert GUT model.

\medskip\noindent\underline{(b) One-loop evolution equations}

The evolution equations for the gauge couplings at one loop can be written
\begin{equation}
{{dg_i}\over {dt}}={{b_ig_i^3}\over {16\pi ^2}} \quad , \quad
{d\over dt}[\alpha_i^{-1}]= -{b_i\over 2\pi}\; ,
\end{equation}
with $t=\ln(\mu /M_Z)$ the logarithmic scale and $\alpha_i=g_i^2/(4\pi)$.
The SM particle content alone gives $b_1=4{1\over 10}$, $b_2=-3{1\over 6}$,
$b_3=-7$.  It is well known that this does not lead to gauge coupling
unification; given $\alpha_2^{-1}$,  $\alpha_1^{-1}$ evolves too fast
compared to $\alpha_3^{-1}$.   Simply adding singlet quarks makes things
worse, however; $\alpha_2^{-1}$ is unchanged, $\alpha_1^{-1}$ evolves
faster and $\alpha_3^{-1}$ evolves more slowly. Some different particle
content is needed.

The MSSM, with additional supersymmetric particle content and
two Higgs doublets, does give successful gauge coupling unification
with the beta functions
\begin{eqnarray}
b_1&=&2n_G+{3\over 5}n_H\;, \\
b_2&=&2n_G+n_H-6\;, \\
b_3&=&2n_G-9\;,
\end{eqnarray}
where $n_G=3$ is the number of light generations of matter and $n_H=1$ is
the number of pairs of light Higgs doublets. In the presence of $n_{x_u}$
up-type
and $n_{x_d}$ down-type singlet quarks, the beta functions are modified
to
\begin{eqnarray}
b_1&=&2n_G+{3\over 5}n_H+{2\over 5}n_{x_d}+{8\over 5}n_{x_u}\;, \\
b_2&=&2n_G+n_H-6\;, \\
b_3&=&2n_G-9+n_{x_d}+n_{x_u}\;.
\end{eqnarray}
These singlet quark contributions upset the MSSM unification as shown
in Fig.~\ref{fig:singlet}. However,
unification can be restored by adding exotic fermions to fill out
one or more representations of SU(5). For example, in an $E_6$ model
the basic ${\bf 27}$ representation has the decompositions ${\bf 16
+ 10 + 1}$ in the $SO(10)$ subgroup, which in turn are ${\bf (10 + 5^*
+1) + (5 + 5^{*\prime}) + 1}$ in $SU(5)$.  If the full ${\bf 27}$ of
fermions is light, the down-type quarks are supplemented by colorless
doublets and singlets, giving
\begin{eqnarray}
b_1&=&3n_G+{3\over 5}n_H\;, \\
b_2&=&3n_G+n_H-6\;, \\
b_3&=&3n_G-9\;.
\end{eqnarray}
where $n_G$ is now the number of light generations of $E_6$ matter,
assuming that the light Higgses are external to the $\bf 27 $ representations.
Thus all $b_i$ are shifted by the same amount $n_G$ from the MSSM case;
all values of $\alpha_i^{-1}(t)$ are shifted down by the same amount
$-b_i t n_G/(2\pi )$ and unification is preserved.  Note incidentally
that with the usual three generations we now have $b_3=0$ and the strong
coupling ceases to run.

However, one usually assumes instead that the pair of Higgs doublets
comes from the ${\bf 27}$, in which case the beta functions are
\begin{eqnarray}
b_1&=&3n_G\;, \\
b_2&=&3n_G-6\;, \\
b_3&=&3n_G-9\;,
\end{eqnarray}
and the gauge coupling unification is again problematic.  One solution
would be to get back to the previous successful beta functions by
adding two new particles with the quantum numbers of two Higgs doublets.
Alternatively, we might have hoped that the two-loop contributions could
rescue gauge coupling unification, since the gauge couplings are all
larger than in the MSSM, making two-loop contributions more important.
Unfortunately the sign of the two-loop term pushes the $SU(3)$ coupling
further away from the electroweak couplings.  Nevertheless, this example
shows that two-loop contributions could be important.

\medskip\noindent\underline{(c) Two-loop evolution equations}

At two-loop level, the evolution equations become
\begin{equation}
{{dg_i}\over {dt}}={g_i\over{16\pi^2 }}\left [b_ig_i^2+{1\over {16\pi^2 }}
\sum _{j=1}^3b_{ij}g_i^2g_j^2
\right ] \;,
\end{equation}
where the one-loop beta functions are
\begin{eqnarray}
b_1&=&{3\over 2}n_{10}+{1\over 2}(n_{5^*}+n_{5^{*\prime }}+n_5)
+{3\over 5}n_H\;, \\
b_2&=&{3\over 2}n_{10}+{1\over 2}(n_{5^*}+n_{5^{*\prime }}+n_5)+n_H-6\;, \\
b_3&=&{3\over 2}n_{10}+{1\over 2}(n_{5^*}+n_{5^{*\prime }}+n_5)-9\; ,
\end{eqnarray}
for an arbitrary number of copies ($n_{10}$, $n_{5^*}$, $n_{5^{*\prime }}$,
$n_5$) in the ${\bf 10}$, ${\bf 5^*}$, ${\bf 5^{*\prime }}$, and ${\bf 5}$
representations of $SU(5)$, and $n_H$ light pairs of Higgs doublets (from
a split representation) .
The two-loop coefficients $b_{ij}$ are listed in Appendix B.
The model-dependent contributions from the Yukawa couplings at two-loop
order have been neglected.   In the absence of split representations
the entries are related by simple $SU(3)$ and $SU(2)$ group factors.
The second ${\bf 5^{*\prime}}$ representation contains the multiplet
$(H, x_d^c)$  which may come from either the ${\bf 10}$ or the
${\bf 16}$ representation of $SO(10)$.

For rank 5 $E_6$ models there is an extra $U(1)$ that enters into the
gauge coupling evolution equations at the two-loop level. There is a
one-parameter family of extra $U(1)$'s orthogonal to $U(1)_Y$. Three
popular models\cite{im} are characterized as follows:

(1) the $SO(10)$ singlet fermion is inert, with respect to the extra
$U(1)$.

(2) the $SU(5)$ singlet fermion in the $\bf 16$ of $SO(10)$ is inert, and

(3) the ${\bf 5^*}$ and ${\bf 5^{*\prime }}$ have exactly the same
quantum numbers.

\noindent
We label the $U(1)$ quantum number of these models by $Y^{\prime }$,
$Y^{\prime \prime }$, and $Y^{\prime \prime \prime }$ respectively, and
list the quantum numbers for the full ${\bf 27}$ of $E_6$ in
Table~\ref{quantum_nos}. Notice that for the first model the ${\bf 16}$ of
$SO(10)$ decomposes as ${\bf 10 + 5^{*\prime } + 1}$. The first two models
could actually arise from an $SO(10)$ theory, since ${\rm Tr}\; Y$ vanishes
across each $SO(10)$ multiplet, while the third model is distinctively $E_6$
(as are all the rest of the rank 5 models).
In the last model it is natural for the entire ${\bf 27}$ to be light.
[We note that the extra abelian groups are often referred to by the
notation $U(1)_\eta $, $U(1)_\chi $ and $U(1)_\psi $\cite{rizzo}; the models
(2) and (3) considered here then correspond to the extra group being
$U(1)_\chi $ and $U(1)_\eta $ respectively, while the model (1) is a
linear combination of $U(1)_\chi $ and $U(1)_\psi $.]
For these three models one obtains, in addition to the RGE coefficients
$b_i$ and $b_{ij}$ already listed, the one-loop coefficients
\begin{eqnarray}
b_p^\prime &=&{1\over 4}n_{10}+{1\over 2}(n_{5^*}+n_5)+{9\over 8}
n_{5^{*\prime }}+{5\over 8}n_N+{13\over 20}n_H\;, \\
b_p^{\prime \prime}&=&{1\over 4}n_{10}+{9\over 8}n_{5^*}+{1\over 2}
(n_{5^{*\prime }}+n_5)+{5\over 8}n_{\nu _L^c}+{2\over 5}n_H\;, \\
b_p^{\prime \prime \prime}&=&{2\over 3}n_{10}+{1\over
12}(n_{5^*}+n_{5^{*\prime }})
+{4\over 3}n_5+{5\over 12}(n_N+n_{\nu _L^c})+{17\over 30}n_H\; ,
\end{eqnarray}
and the two-loop contributions are listed in the Appendix B.
We use the subscript $p$ (for ``prime'') to distinguish
the $U(1)'$, $U(1)''$ and $U(1)'''$ gauge couplings and their RGE coefficients.
Here $n_N$ and $n_{\nu _L^c}$ are the number of light singlets with the
quantum numbers given in Table~\ref{quantum_nos}.
In a general model with more than one $U(1)$ factor, one must account for
the mixing between the $U(1)$'s in the renormalization group equations
\cite{dacq}. This complication does not
arise if one considers only unification trajectories.
In any event the practical
effect of the extra $U(1)$ on the evolution is small.
The above equations have been derived for an arbitrary number of
different representations of $SU(5)$, but  split representations in the ${\bf
10}$ of $SO(10)$ have been allowed for
(${\bf 10}\to {\bf 5}+{\bf 5^*}\to H+\overline{H}$),
since they may be needed to
achieve gauge coupling unification.   The beta functions for
$U(1)^{\prime }$ and $U(1)^{\prime \prime }$ are related
by $n_{5^*}\leftrightarrow n_{5^{*\prime}}$ and
$n_N\leftrightarrow n_{\nu _L^c}$.
One can also consider the continuous family of rank 5 $E_6$ models that
include the three above, but as far as gauge coupling unifications is concerned
they offer no new features.

An $E_6$ model with three light generations would have
$n_{10}=n_{5^*}=n_{\nu _L^c}=n_5=n_{5^{*\prime }}=n_N=3$
(from the usual decomposition of the ${\bf 27}$ representation).
If only complete ${\bf 27}$ multiplets of $E_6$ occur,
the above coefficients become universal, namely
\begin{eqnarray}
b_{1}&=&b_p^\prime =b_p^{\prime \prime }=b_p^{\prime \prime \prime }
=3n_G\;, \\
b_{2}&=&3n_G-6\;, \\
b_{3}&=&3n_G-9\;,
\end{eqnarray}
and
\begin{equation}
b_{ij}^{\prime } = b_{ij}^{\prime \prime } =b_{ij}^{\prime \prime \prime }
= \left( \begin{array}
{c@{\quad}c@{\quad}c@{\quad}c}
3 & 1 & 3  & 8 \\
1 & 3 & 3  & 8 \\
1 & 1 & 21 & 8 \\
1 & 1 & 3  & 34
\end{array} \right)n_G+\left( \begin{array}
{c@{\quad}c@{\quad}c@{\quad}c}
0 &   &     &    \\
  & 0 &     &    \\
  &   & -24 &    \\
  &   &     & -54
\end{array} \right) \;.
\end{equation}

\medskip\noindent\underline{(d) Specific scenarios}

These results allow us to examine gauge unification with different
numbers of light generations of exotic matter, with or without a pair of
light Higgses from a split representation. We recall that the MSSM
has $n_{10}=n_{5^*}=3$, $n_5=n_{5^{*\prime }}=0$  with $n_H=1$
(the distinction between the ${\bf 5^*}$ and ${\bf 5^{*\prime }}$ is
immaterial as far as $SU(5)$ is concerned), and the Higgs
contribution is vital for successful unification.\\
(i) Extensions of the MSSM.  Adding just one or two generations of
$SO(10)$ ${\bf 10}$-plet matter ($n_5=n_{5^{*\prime }}=1$ and keeping
$n_H=1$) yields successful gauge coupling unification as shown in
Fig.~\ref{fig:oneg1h}. With three generations, however, the two-loop
corrections threaten to spoil unification; they also make $\alpha_3$
increase toward the GUT scale (although $b_3=0$ at one-loop).
On the other hand, one
expects the low-energy threshold corrections to be more significant in this
case\cite{ralf}, and ultimately the success of unification depends on
the details of  the low-energy spectrum. [Ref.~\cite{ralf} assumes that the
$SU(5)$ multiplets are degenerate at the GUT scale.]\\
(ii) $E_6$ based models.  Here one needs some split representation
since otherwise the electroweak couplings do not run fast enough for
successful unification. (Some attempts at $E_6$ phenomenology have
assumed that the Higgs pair comes from a complete light representation.)
The two-loop contributions do not help since they tend to slow the
running of the strong coupling constant, or even make it grow in
the case where the one-loop beta function $b_3$ is exactly zero
(as happens for three generations of light $E_6$ matter). Although
asymptotic freedom is lost above the exotic fermion mass scale, this
is not necessarily a problem for gauge coupling unification as long as
the two-loop effects do not make $\alpha_3$ nonperturbative below the
GUT scale. One has gauge unification at the same scale as in the MSSM
(neglecting threshold correction), with unification coupling
$\alpha _3(M_{\rm exotic})$ still perturbative, though significantly
larger than in the MSSM.  \\
(iii) $E_6$ models with $n_G=3$.  In all $E_6$ models one expects
$\alpha_3$ to run more slowly than in the MSSM due to the extra matter
in the $\bf 5$ and $\bf 5^{*\prime }$ representations, keeping $\alpha_3$
bigger and making two-loop contributions more important.   We find that
the latter destroy unification if there are three light generations of
$E_6$ matter (very similar to the case in
Fig.~\ref{fig:oneg1h}) even when an extra Higgs pair is included;
there are model-dependent deviations from the curves in
Fig.~\ref{fig:oneg1h}, due to the presence of the extra U(1), but
these are very small.     Gauge coupling unification would require
$\alpha _3(M_Z)$ to be reduced below the MSSM prediction by about 15\%.
Unification could conceivably still be rescued by threshold corrections
from large splittings in the $SU(5)$ multiplets at low energy.
With three complete generations excluding the Higgses, the situation is
much worse as shown in Fig.~\ref{fig:threeg}.  \\
(iv) $SU(5)$ models with an extra $\bf 10$ and $\bf 10^*$. This case is
similar to case (iii), since the beta function coefficients $b_3$ and $b_{33}$
are the same. Two-loop contributions to the RGE's become relatively more
important, and gauge coupling unification becomes problematic without
large threshold corrections.

We have followed a philosophy of preferring the least number of split
representations possible. It is possible to make gauge coupling unification
work without an intermediate scale far removed from the electroweak scale
by relaxing this constraint. In fact, a nonsupersymmetric left-right $E_6$
model has been proposed recently\cite{ma2} in which the $SU(2)_R$ is broken
at 1 TeV.

\medskip\noindent\underline{(e) Yukawa evolution}

One can consider the evolution of Yukawa couplings in this new
scenario where the QCD coupling does not run as rapidly as it does in the
MSSM. For $b_3=0$ one can immediately solve the one-loop renormalization group
equations (neglecting the $SU(2)$ and $U(1)$ couplings which are small except
near the GUT scale).

Consider the superpotential
\begin{eqnarray}
W&=&\lambda _t H_{2,3}Qt^c+\lambda _b H_{1,3}Qb^c
+\lambda _{\tau } H_{1,3}L\tau ^c
+\lambda _{S_i} S_3H_{1,i}H_{2,i} + \lambda _{d_i} S_3x_{d_i}x_{d_i}^c\;.
\end{eqnarray}
We define $H_{1,3}$, $H_{2,3}$ and $S_3$ to be the linear combination of the
Higgs doublets and singlets that acquire a vev.
For the top Yukawa
coupling one has (assuming that $\lambda _b$ and $\lambda _{\tau }$ can be
neglected)

\begin{eqnarray}
{{d\lambda _t}\over {dt}}&=&{{\lambda _t}\over {16\pi^2}}\left [
-\sum _i c_ig_i^2-3g_2^2-{16\over 3}g_3^2+6\lambda _t^2
+\lambda _{S_3}^2
\right ]\;,
\end{eqnarray}
where $g_i$ are the running $U(1)$ gauge couplings (and hence contain some
model dependence).
The new couplings that arise from the presence of an electroweak Higgs singlet
$S$ evolve as
\begin{eqnarray}
{{d\lambda _{S_3}}\over {dt}}&=&{{\lambda _{S_3}}\over {16\pi^2}}\left [
-\sum _i d_ig_i^2-3g_2^2+3\lambda _t^2+4\lambda _{S_3}^2
+3\sum _j\lambda _{d_j}^2\right ]\;, \\
{{d\lambda _{S_i}}\over {dt}}&=&{{\lambda _{S_i}}\over {16\pi^2}}\left [
-\sum _i d_ig_i^2-3g_2^2+3\lambda _t^2+4\lambda _{S_i}^2
\right ]\quad i\ne 3\;, \\
{{d\lambda _{d_i}}\over {dt}}&=&{{\lambda _{d_i}}\over {16\pi^2}}\left [
-\sum _i e_ig_i^2-{16\over 3}g_3^2+2\lambda _s^2
+2\lambda _{d_i}^2+3\sum _j\lambda _{d_j}^2\right ]\;,
\end{eqnarray}
where $c_i$, $d_i$ and $e_i$ are some (in general model-dependent) coefficients
of the $U(1)$ gauge couplings. We do not assume any $SU(5)$ relation between
$\lambda _{S_i}$ and $\lambda _{d_i}$.

Since the gauge coupling values near the GUT scale are much larger
than they are in the minimal supersymmetric model, one expects the
top Yukawa coupling to be driven much faster to its fixed point value
from below. The general result is that
if singlet quarks are accompanied by other exotics fermions to fill out
representations of SU(5), then the top-quark is driven to its fixed point
value ($\lambda _t^2\approx 8/9g_3^2$) over a wide range of values for the
top quark GUT scale Yukawa; see Fig.~\ref{fig:yukawa}.
The presence of the coupling $\lambda _S$ in the
top quark coupling RGE could soften the attraction to the fixed point, but
requiring it to remain perturbative up to the GUT scale prevents it from
destroying the fixed point solution, as shown in the NMSSM model\cite{nmssm}.

The linear combinations $H_{1,3}$, $H_{2,3}$ and $S_3$ of Higgs
fields acquire vevs $v_2$, $v_1$ and $v_s$ and one defines
$\tan \beta =v_2/v_1$. One gets the usual relations that one has in the
MSSM model
\begin{eqnarray}\label{lambda_b}
\lambda_b(m_t) = {\sqrt2\, m_b(m_b)\over\eta_b v\cos\beta}\,, \qquad
\lambda_\tau(m_t) = {\sqrt2m_\tau(m_\tau)\over \eta_\tau v\cos\beta}\,, \qquad
\lambda_t(m_t) = {\sqrt2 m_t(m_t)\over v\sin\beta}\;,
\end{eqnarray}
in addition to mass relations for the squark singlets and exotic leptons
\begin{eqnarray}
\lambda_{d_i}={{m_{x_{d_i}}}\over v_s}\,, \qquad
\lambda_{S_i}={{m_{H_i}}\over v_s}\;,
\end{eqnarray}
ignoring mixing.

The singlet quark Yukawa couplings $\lambda _{d_i}$ also have infrared
fixed points\cite{drees}, essentially given by the condition
\begin{eqnarray}
2\lambda _{d_3}^2+3\sum _j\lambda _{d_j}^2={16\over 3}g_3^2\;.
\end{eqnarray}
Unfortunately this does not yield a prediction for the singlet quark mass since
the Higgs singlet vev is a priori unknown. At best one can obtain
an upper limit on the ratio $m_{x_d}/M_{Z^\prime }$\cite{drees}. When the
singlet quark Yukawa is at its fixed point, it saturates this upper limit.

One can also consider the implications of bottom-tau unification in the
context of these $E_6$ models. The evolution of the Yukawa couplings is the
same as it is for the MSSM,
\begin{eqnarray}
{{dR_{b/\tau}}\over {dt}}&=&{{R_{b/\tau}}\over {16\pi^2}}\left [
-{4\over 3}g_1^2-{16\over 3}g_3^2+\lambda _t^2+3\lambda _b^2
-3\lambda _{\tau }^2\right ]
\nonumber \\ \nonumber
\end{eqnarray}
where $R_{b/\tau}\equiv {{\lambda _b}\over {\lambda _{\tau}}}$.
However, we find that since the gauge couplings are larger over the entire
range of scales between $M_{GUT}$ and the electroweak scale, the Yukawa
couplings have to be correspondingly larger to cancel off the contributions
from the gauge couplings. In the MSSM, the top Yukawa is often forced into
the infrared fixed point region. In the $E_6$ model with three light
generations, we find that there is
no solution that gives an acceptable value for $m_b$\cite{ralf}.

\section{Conclusions}

Quark singlets offer an interesting example of physics beyond the SM.
They mix with the ordinary fermions.  They impact a wide variety of
experimental measurements, as they generate tree-level FCNC's,
introduce unitarity violation in the SM CKM matrix, influence
neutral meson-antimeson oscillations, and modify CP asymmetries.
These objects can be produced by strong, electromagnetic and
weak-neutral-current interactions, and produce interesting decay
signatures.  Their masses must generally exceed ${1\over 2}M_Z$;
higher limits 85-131 GeV apply in certain particular scenarios
(see Section IV).

We have collected the available bounds on singlet quark mixing;
some have been updated; some, such as the
$B^0,D^0\to\mu^+\mu^-$ and $D\to\mu^+\mu^-X$ constraints
and the unitarity implications of FDNC bounds, have not appeared
explicitly before (see Sections III-VI).  The present limits on the
$4\times 4$ mixing matrix elements connecting one new singlet quark
to standard quarks may be summarized as follows.
$$\begin{array}{ccll}
\underline{Q=-{1\over 3}\; {\rm case}}&& \underline{{\rm limit}}
& \underline{{\rm origin}}        \\
|V_{od}|            & \alt & 0.048             & {\rm global\; FDNC}   \\
|V_{os}|            & \alt & 0.060             & {\rm global\; FDNC}   \\
|V_{ob}|            & \alt & 0.045             & {\rm global\; FDNC}   \\
|V_{ux}|            & \alt & 0.08              & {\rm CKM+unitarity}\\
|V_{cx}|            & \alt & 0.09              & {\rm FDNC+unitarity}\\
|V_{tx}|            & \alt & 0.09              & {\rm FDNC+unitarity}\\
|V_{ox}|            & \agt & 0.996             & {\rm FDNC+unitarity}\\
|V_{os}||V_{od}|    & \alt & 3\times 10^{-4}
                                      & \epsilon ,\delta m_K({\rm tree})\\
|V_{ob}||V_{od}|    & \alt & 8\times 10^{-4}   & \delta m_B({\rm tree})\\
|V_{ob}||V_{os}|    & \alt & 2\times 10^{-3}   & B\to\ell^+\ell^-X\\
|V_{cx}||V_{ux}|    & \alt & (1.3 {\rm GeV})/m_x      & \delta m_D({\rm box})\\
|Re(V^*_{od}V_{os})||Im(V^*_{od}V_{os})|
                    &\alt& 3\times 10^{-10}    & \epsilon_K \\
|Re(V_{od}^*V_{os})| &\alt& 7\times 10^{-6}     & K_L\to\mu\mu
\end{array}$$
$$\begin{array}{ccll}
\underline{Q= {2\over 3}\; {\rm case}}&& \underline{{\rm limit}}
       & \underline{{\rm origin}}    \\
|\hat V_{uo}|       & \alt & 0.049             & {\rm global\; FDNC}\\
|\hat V_{co}|       & \alt & 0.065             & {\rm global\; FDNC}\\
|\hat V_{to}|       & \alt & 1.0               & {\rm unitarity}\\
|\hat V_{xd}|       & \alt & 0.15              & {\rm CKM+unitarity}\\
|\hat V_{xs}|       & \alt & 0.56              & {\rm CKM+unitarity}\\
|\hat V_{xb}|       & \alt & 1.0               & {\rm unitarity}\\
|\hat V_{co}||\hat V_{uo}|&\alt& 9\times 10^{-4}& \delta m_D({\rm tree})\\
|\hat V_{xd}||\hat V_{xb}|&\alt& 0.03          & {\rm CKM+unitarity}
\end{array}$$

We have discussed possible effects of singlet quarks on
$b\to d\gamma ,s\gamma$ decays, and have illustrated how a $u$-type
singlet could either increase or decrease the SM rate for
$b\to s\gamma$ (Fig.\ref{fig:bsgam}).  We have pointed out that
small $x-q$ mixing  reduces the branching fraction for
$Z\to q\bar q$ decays; in the interesting case $Z\to b\bar b$, this
would worsen the present discrepancy between SM and experiment.
We have given new illustrations of ways that singlet quarks can cause
substantial deviations from SM expectations for CP-asymmetries of
neutral $B$ decays (Figs.\ref{fig:beta}-\ref{fig:alpha}).
The asymmetry ${\rm Im}\; \lambda (B_d\to \psi K_S)$ can have the
opposite sign to the SM value.

In the GUT context, singlet quarks cannot simply be added by themselves
to the SM or MSSM, since this would destroy gauge coupling unification
(Fig.\ref{fig:singlet}); they must be accompanied by other members
of exotic fermion multiplets.  Down-type singlet quarks are readily
accommodated in grand unified extensions of the SM;  as a minimal
scenario, they can be realized by
adding one or more extra generations of $\bf 5$ and $\bf 5^*$
representations of $SU(5)$, that imply extra vector-doublet leptons
too.  This exotic matter together with the SM matter content fits
into the $\bf 27$ representation of $E_6$ (which decomposes to
$\bf 10+\bf 5^*+\bf 1+\bf 5+\bf 5^{*\prime}+\bf 1$ in an $SU(5)$ subgroup).
Adding extra complete multiplets of $SU(5)$ preserves (at the one-loop level)
the successful unification of gauge couplings in the MSSM,
since a complete multiplet contributes equally to the evolution
of each coupling.  However, more than three generations of exotic matter
will destroy asymptotic freedom for $\alpha_3$ at one loop
(Fig.\ref{fig:oneg1h}).

As in the MSSM, gauge coupling unification in a desert model can be achieved
by assuming that split representations exist. In the context of models with
singlet quarks, this means that there must be an additional pair of light
Higgs doublets, in addition to the pairs that are included
with the singlet quarks in the  $\bf 5$ and $\bf 5^*$ representations.
(The MSSM is then a special case consisting of no singlet quarks and one light
pair of Higgs doublets).

An up-type singlet quark is not contained as elegantly in GUT Models; it
does not appear in the smallest representations, and its role is less clear. As
a minimal prescription, it can be introduced by adding one extra light
$\bf 10$ and one $\bf 10^*$  representation of $SU(5)$ that get their mass
from an $SU(5)$ singlet Higgs boson; this implies extra vector-doublet
quarks and a vector-singlet charged lepton too, preserving MSSM gauge
coupling unification with $b_3=0$ at one loop (Fig.\ref{fig:oneg1h}).
Less minimally, it can also be realized in the $SO(10)$ group with an
extra light $\bf 45 $ (adjoint) representation (which decomposes to
$\bf 24+\bf 10^*+\bf 10+\bf 1$ in an $SU(5)$ subgroup), but this leads to
nonperturbative gauge couplings at the GUT scale if the entire $\bf 45$
is required to be light.

Two-loop effects are typically small in most GUT models, but if one includes
extra representations of matter then the evolution of the strong coupling
is diminished and it might even increase (no asymptotic freedom) toward the
GUT scale. In a situation where the strong gauge coupling does not evolve at
the one-loop level, we find that the two-loop effects become relatively more
important and can make gauge coupling unification problematic, e.g. for
three complete generations of $E_6$ ${\bf 27}$-plet matter.  However one
expects the low-energy threshold corrections to be more significant in this
case, and ultimately the success of unification depends on
the details of  the low-energy spectrum. Two-loop effects also threaten
perturbativity and asymptotic freedom of $\alpha_3$
(Figs.\ref{fig:oneg1h}-\ref{fig:threeg}).

Fixed points play a role in these extended model, with the top quark and
the down-type singlet(s) masses possibly determined by the gauge couplings
and the associated vevs.  With extended matter content and larger gauge
couplings, the top Yukawa coupling is driven to its fixed point faster
than before (Fig.\ref{fig:yukawa}).  However, the Yukawa unification
condition $\lambda_b(M_G^{})=\lambda_\tau (M_G^{})$ becomes harder to
accomodate, and fails in the $E_6$ model with three light generations.

\section{Appendix A}
In this appendix we collect a few results needed for the analysis of
$b\to q\gamma$ ($q=s,d$) in Section VI.
The SM magnetic and chromomagnetic couplings for flavor-changing
$b\to q$ decays  via $W$ loops are given by\cite{bg}
\begin{eqnarray}
f_{\gamma }^{(1)}&=&{{7-5x-8x^2}\over {36(x-1)^3}}+{{x(3x-2)}\over
{6(x-1)^4}}
\ln x\;, \\
f_g^{(1)}&=&{{2+5x-x^2}\over {12(x-1)^3}}-{{x}\over {2(x-1)^4}}
\ln x\;,
\end{eqnarray}
where $x=m_t^2/M_W^2$. A down-type singlet quark induces additional
$Z$ loops, giving the replacements
\begin{eqnarray}
{3\over 2} xf_{\gamma }^{(1)}&\rightarrow & {3\over 2} xf_{\gamma }^{(1)}
+\left ({{z_{qb}}\over {V_{tb}V_{tq}^*}}\right )
\left ({19\over 54}-{2\over 81}\sin ^2\theta _W\right )\;, \\
{3\over 2} xf_g^{(1)}&\rightarrow & {3\over 2} xf_g^{(1)}
+\left ({{z_{qb}}\over {V_{tb}V_{tq}^*}}\right )
\left ({4\over 9}+{2\over 27}\sin ^2\theta _W\right )\;.
\end{eqnarray}
These substitutions then enter into the values of the Wilson coefficients
where
\begin{eqnarray}
   c_7(M_W) = \left [{3\over 2} xf_{\gamma}^{(1)}(x)
+\left ({{z_{qb}}\over {V_{tb}V_{tq}^*}}\right )
\left ({19\over 54}-{2\over 81}\sin ^2\theta _W\right )
\right ]
\;, \label{c7II} \\
   c_8(M_W) = \left [{3\over 2} xf_g^{(1)}(x)
+\left ({{z_{qb}}\over {V_{tb}V_{tq}^*}}\right )
\left ({4\over 9}+{2\over 27}\sin ^2\theta _W\right )
\right ]
\;. \label{c8II}
\end{eqnarray}
The coefficients from the $8\times 8$ anomalous dimension matrix are\cite{qcd}
\begin{equation}
\begin{array}{ccccccccc}
\vspace{0.2cm}
a_i = ( &    \frac{14}{23},     &     \frac{16}{23},   &
\frac{6}{23},&-\frac{12}{23},
        &      0.4086,       &      -0.4230,     & -0.8994,  &   0.1456     )
\\
h_i = ( &\frac{626126}{272277}, &-\frac{56281}{51730}, &-\frac{3}{7}, &
-\frac{1}{14},
        &     -0.6494,       &      -0.0380,     & -0.0186,  &  -0.0057     )
\end{array}
\end{equation}
In the case of up-type singlet quarks the Wilson coefficients are
\begin{eqnarray}
   c_7^t(M_W) &=& {3\over 2} xf_{\gamma}^{(1)}(x)\;, \\
   c_7^x(M_W) &=& {3\over 2} yf_{\gamma}^{(1)}(y)\;, \\
   c_8^t(M_W) &=& {3\over 2} xf_g^{(1)}(x)\;, \\
   c_8^x(M_W) &=& {3\over 2} yf_g^{(1)}(y)\;,
\end{eqnarray}
where $y=m_{x_u}^2/M_W^2$.

\section{Appendix B}
We here collect some two-loop results needed in Section VIII.
The two-loop RGE coefficients for an arbitrary number of ${\bf 10}$,
${\bf 5^*}$, and ${\bf 5}$ representations are
\begin{mathletters}
\begin{eqnarray}
b_{11}&=&{23\over 10}n_{10}+{7\over 30}(n_{5^*}+n_{5^{*\prime }}+n_5)
+{9\over 25}n_H \;, \\
b_{12}&=&{3\over 10}n_{10}+{9\over 10}(n_{5^*}+n_{5^{*\prime }}+n_5)
+{9\over 5}n_H \;, \\
b_{13}&=&{24\over 5}n_{10}+{16\over 15}(n_{5^*}+n_{5^{*\prime }}+n_5)\;, \\
b_{21}&=&{1\over 10}n_{10}+{3\over 10}(n_{5^*}+n_{5^{*\prime }}+n_5)
+{3\over 5}n_H \;, \\
b_{22}&=&{21\over 2}n_{10}+{7\over 2}(n_{5^*}+n_{5^{*\prime }}+n_5)+7n_H-24
\;, \\
b_{23}&=&8n_{10} \;, \\
b_{31}&=&{3\over 5}n_{10}+{2\over 15}(n_{5^*}+n_{5^{*\prime }}+n_5) \;, \\
b_{32}&=&3n_{10} \;, \\
b_{33}&=&17n_{10}+{17\over 3}(n_{5^*}+n_{5^{*\prime }}+n_5)-54 \;,
\end{eqnarray}
\end{mathletters}
in the $g_1$, $g_2$, $g_3$ basis.

The two-loop coefficients for the $E_6$ models are
\begin{mathletters}
\begin{eqnarray}
b_{pp}^{\prime }&=&{1\over 40}n_{10}+{1\over 5}
(n_{5^*}+n_5)+{81\over 80}n_{5^{*\prime }}
+{25\over 16}n_N+{97\over 200}n_H \;, \\
b_{p2}^{\prime }&=&{3\over 20}n_{10}+{1\over 5}
(n_{5^*}+n_5)+{9\over 20}n_{5^{*\prime }}
+{39\over 100}n_H \;, \\
b_{p3}^{\prime }&=&{9\over 20}n_{10}+{3\over 5}(n_{5^*}+n_5)
+{27\over 20}n_{5^{*\prime }}+{39\over 20}n_H \;, \\
b_{p4}^{\prime }&=&{6\over 5}n_{10}+{8\over 5}(n_{5^*}
+n_5)+{18\over 5}n_{5^{*\prime }} \;, \\
b_{1p}^{\prime }&=&{3\over 20}n_{10}+{1\over 5}
(n_{5^*}+n_5)+{9\over 20}n_{5^{*\prime }}
+{39\over 100}n_H \;, \\
b_{11}^{\prime }&=&{23\over 10}n_{10}+{7\over 30}
(n_{5^*}+n_{5^{*\prime }}+n_5)+{9\over 25}n_H \;, \\
b_{12}^{\prime }&=&{3\over 10}n_{10}+{9\over 10}
(n_{5^*}+n_{5^{*\prime }}+n_5)+{9\over 5}n_H \;, \\
b_{13}^{\prime }&=&{24\over 5}n_{10}+{16\over 15}
(n_{5^*}+n_{5^{*\prime }}+n_5)\;, \\
b_{2p}^{\prime }&=&{3\over 20}n_{10}+{1\over 5}
(n_{5^*}+n_5)+{9\over 20}n_{5^{*\prime }}
+{13\over 20}n_H \;, \\
b_{21}^{\prime }&=&{1\over 10}n_{10}+{3\over 10}
(n_{5^*}+n_{5^{*\prime }}+n_5)+{3\over 5}n_H \;, \\
b_{22}^{\prime }&=&{21\over 2}n_{10}+{7\over 2}
(n_{5^*}+n_{5^{*\prime }}+n_5)-24 \;, \\
b_{23}^{\prime }&=&8n_{10} \;, \\
b_{3p}^{\prime }&=&{3\over 20}n_{10}+{1\over 5}
(n_{5^*}+n_5)+{9\over 20}n_{5^{*\prime }} \;, \\
b_{31}^{\prime }&=&{3\over 5}n_{10}+{2\over 15}
(n_{5^*}+n_{5^{*\prime }}+n_5) \;, \\
b_{32}^{\prime }&=&3n_{10} \;, \\
b_{33}^{\prime }&=&17n_{10}+{17\over 3}(n_{5^*}+n_{5^{*\prime }}+n_5)-54 \;,
\end{eqnarray}
\end{mathletters}
\begin{mathletters}
\begin{eqnarray}
b_{pp}^{\prime \prime}&=&{1\over 40}n_{10}+{81\over 80}n_{5^*}+{1\over
5}(n_{5^{*\prime }}+n_5)
+{25\over 16}n_{\nu _L^c}+{97\over 200}n_H \;, \\
b_{p1}^{\prime \prime}&=&{3\over 20}n_{10}
+{9\over 20}n_{5^*}+{1\over 5}(n_{5^{*\prime }}+n_5)+{6\over 25}n_H \;, \\
b_{p2}^{\prime \prime}&=&{9\over 20}n_{10}+{27\over 20}n_{5^*}
+{3\over 5}(n_{5^{*\prime }}+n_5)
+{39\over 20}n_H \;, \\
b_{p3}^{\prime \prime}&=&{6\over 5}n_{10}+{18\over 5}n_{5^*}+{8\over 5}
(n_{5^{*\prime }}+n_5) \;, \\
b_{p4}^{\prime \prime}&=&{3\over 20}n_{10}+{9\over 20}n_{5^*}+{1\over 5}
(n_{5^{*\prime }}+n_5)
+{6\over 25}n_H \;, \\
b_{11}^{\prime \prime}&=&{23\over 10}n_{10}+{7\over 30}
(n_{5^*}+n_{5^{*\prime }}+n_5)+{9\over 25}n_H \;, \\
b_{12}^{\prime \prime}&=&{3\over 10}n_{10}+{9\over 10}
(n_{5^*}+n_{5^{*\prime }}+n_5)+{9\over 5}n_H \;, \\
b_{13}^{\prime \prime}&=&{24\over 5}n_{10}+{16\over 15}
(n_{5^*}+n_{5^{*\prime }}+n_5)\;, \\
b_{2p}^{\prime \prime}&=&{3\over 20}n_{10}+{9\over 20}n_{5^*}
+{1\over 5}(n_{5^{*\prime }}+n_5)
+{13\over 20}n_H \;, \\
b_{21}^{\prime \prime}&=&{1\over 10}n_{10}+{3\over 10}
(n_{5^*}+n_{5^{*\prime }}+n_5)+{3\over 5}n_H \;, \\
b_{22}^{\prime \prime}&=&{21\over 2}n_{10}+{7\over 2}
(n_{5^*}+n_{5^{*\prime }}+n_5)-24 \;, \\
b_{23}^{\prime \prime}&=&8n_{10} \;, \\
b_{3p}^{\prime \prime}&=&{3\over 20}n_{10}+{9\over 20}n_{5^*}
+{1\over 5}(n_{5^{*\prime }}+n_5) \;, \\
b_{31}^{\prime \prime}&=&{3\over 5}n_{10}+{2\over 15}
(n_{5^*}+n_{5^{*\prime }}+n_5) \;, \\
b_{32}^{\prime \prime}&=&3n_{10} \;, \\
b_{33}^{\prime \prime}&=&17n_{10}+{17\over 3}(n_{5^*}+n_{5^{*\prime }}+n_5)
-54 \;,
\end{eqnarray}
\end{mathletters}
\begin{mathletters}
\begin{eqnarray}
b_{pp}^{\prime \prime \prime}&=&{8\over 45}n_{10}+{1\over
180}(n_{5^*}+n_{5^{*\prime }})+{64\over 45}n_5
+{25\over 36}(n_N+n_{\nu _L^c})+{257\over 450}n_H \;, \\
b_{p1}^{\prime \prime \prime}&=&{2\over 5}n_{10}
+{1\over 30}(n_{5^*}+n_{5^{*\prime }})+{8\over 15}n_5+{17\over 50}n_H \;, \\
b_{p2}^{\prime \prime \prime}&=&{6\over 5}n_{10}+{1\over 10}(n_{5^*}
+n_{5^{*\prime }})+{8\over 5}n_5
+{17\over 10}n_H \;, \\
b_{p3}^{\prime \prime \prime}&=&{16\over 5}n_{10}+{4\over 15}(n_{5^*}
+n_{5^{*\prime }})
+{64\over 15}n_5 \;, \\
b_{1p}^{\prime \prime \prime}&=&{2\over 5}n_{10}+{1\over 30}(n_{5^*}
+n_{5^{*\prime }})+{8\over 15}n_5
+{17\over 50}n_H \;, \\
b_{11}^{\prime \prime \prime}&=&{23\over 10}n_{10}+{7\over 30}(n_{5^*}
+n_{5^{*\prime }}+n_5)+{9\over 25}n_H \;, \\
b_{12}^{\prime \prime \prime}&=&{3\over 10}n_{10}+{9\over 10}(n_{5^*}
+n_{5^{*\prime }}+n_5)+{9\over 5}n_H \;, \\
b_{13}^{\prime \prime \prime}&=&{24\over 5}n_{10}+{16\over 15}(n_{5^*}
+n_{5^{*\prime }}+n_5)\;, \\
b_{2p}^{\prime \prime \prime}&=&{2\over 5}n_{10}+{1\over 30}(n_{5^*}
+n_{5^{*\prime }})+{8\over 15}n_5
+{17\over 30}n_H \;, \\
b_{21}^{\prime \prime \prime}&=&{1\over 10}n_{10}+{3\over 10}(n_{5^*}
+n_{5^{*\prime }}+n_5)+{3\over 5}n_H \;, \\
b_{22}^{\prime \prime \prime}&=&{21\over 2}n_{10}+{7\over 2}(n_{5^*}
+n_{5^{*\prime }}+n_5)-24 \;, \\
b_{23}^{\prime \prime \prime}&=&8n_{10} \;, \\
b_{3p}^{\prime \prime \prime}&=&{2\over 5}n_{10}+{1\over 30}
(n_{5^*}+n_{5^{*\prime }})+{8\over 15}n_5 \;, \\
b_{31}^{\prime \prime \prime}&=&{3\over 5}n_{10}+{2\over 15}
(n_{5^*}+n_{5^{*\prime }}+n_5) \;, \\
b_{32}^{\prime \prime \prime}&=&3n_{10} \;, \\
b_{33}^{\prime \prime \prime}&=&17n_{10}+{17\over 3}(n_{5^*}
+n_{5^{*\prime }}+n_5)-54 \;.
\end{eqnarray}
\end{mathletters}

\newpage

\begin{table}
\caption{\label{mixing} FCNC effects of singlet quarks}
\smallskip
\begin{tabular}{l|cc}
& $Q_x=-\case1/3$& $Q_x=\case2/3$\\ \hline
$K^0$-$\bar K_0$ osc.& $Z$-exchange& box\\
$D^0$-$\bar D^0$ osc.& box& $Z$-exchange\\
$B_d^0$-$\bar B_d^0$ osc.& $Z$-exchange& box\\
$B_s^0$-$\bar B_s^0$ osc.& $Z$-exchange& box
\end{tabular}
\end{table}

\begin{table}
\caption{\label{quantum_nos} Quantum numbers of rank 5 $E_6$ models}
\smallskip
\begin{tabular}{l|ccr|cc|c|cc|cc|c}
 & & ${\bf 10}$ & & ${\bf 5^*}$ & & ${\bf 1}$ & ${\bf 5^{*\prime }}$ & &
${\bf 5}$ & &
{\bf 1}\\ \hline
Model & $Q_L$ & $u_L^c$ & $e_L^c$ & $d_L^c$ & $L$ & $\nu _L^c$ &
$H$ & $x_d^c $ & $\overline{H}$ & $x_d$ & N\\ \hline
$Y\left (\times \sqrt{3\over 5}\right )$ & ${1\over 6}$ & $-{2\over 3}$ & 1
&${1\over 3}$ & $-{1\over 2}$ & 0 & $-{1\over 2}$ & ${1\over 3}$ &
${1\over 2}$ &
$-{1\over 3}$ & 0 \\
$Y^\prime \left (\times \sqrt{1\over 40}\right )$ & $-1$ & $-1$ & $-1$ &
$-2$ & $-2$ & 0 & 3 & 3 &
2 & 2 & $-5$\\
$Y^{\prime \prime }\left (\times \sqrt{1\over 40}\right )$ &
$-1$ & $-1$ & $-1$ &
$3$ & $3$ & $-5$  & $-2$ & $-2$ &
2 & 2 & 0\\
$Y^{\prime \prime \prime }\left (\times \sqrt{1\over 15}\right )$ &
$-1$ & $-1$ & $-1$ &
${1\over 2}$ & ${1\over 2}$ & $-{5\over 2}$ & ${1\over 2}$ & ${1\over 2}$ &
2 & 2 & $-{5\over 2}$
\end{tabular}
\end{table}

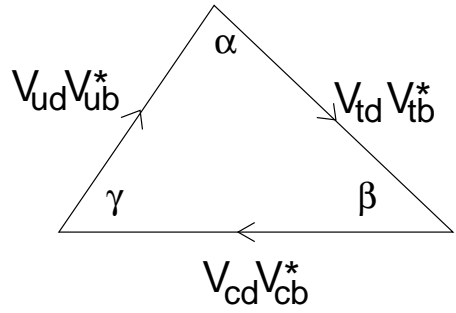
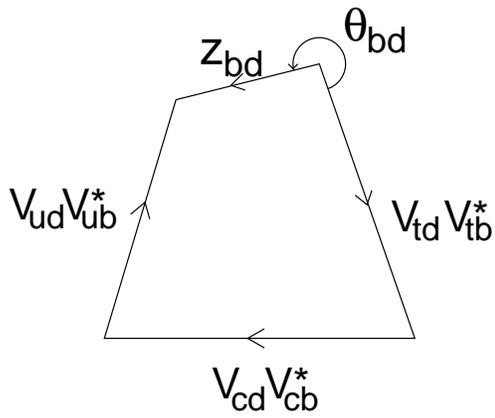
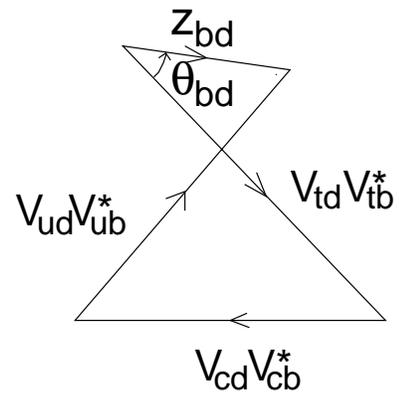
\begin{figure}
\caption{\label{fig:quad} Unitarity of the $3\times 3$ CKM matrix
implies triangle relations like the one shown. In $4\times 4$
mixing cases the triangle relations become quadrangle relations.}
\end{figure}
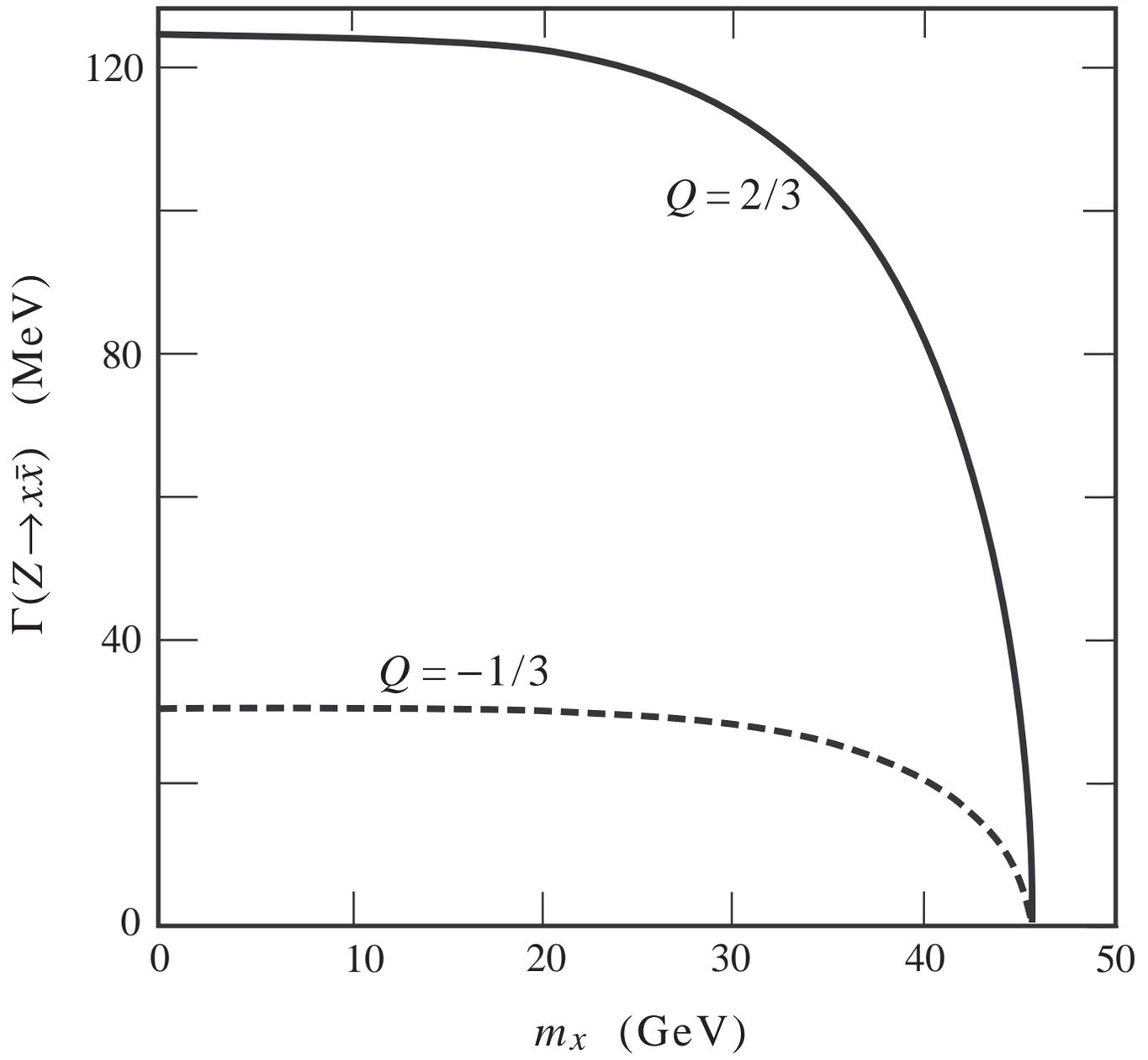
\begin{figure}
\caption{\label{fig:xx} The $Z\to \bar xx$ contribution to the toal $Z$ decay
width $\Gamma_Z$ is shown versus $m_x$ for $Q_x=-{1\over 3}$ and ${2\over 3}$.}
\end{figure}
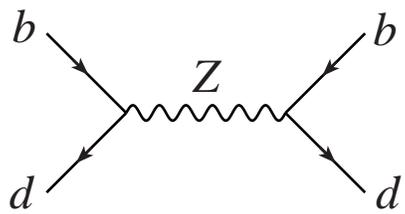
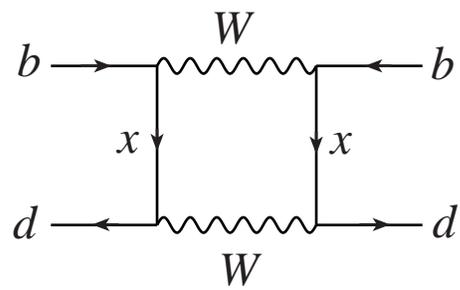
\begin{figure}
\caption{\label{fig:treebox} Singlet quark mixing can give meson-antimeson
oscillations via induced FCNC tree diagrams (a) and via box diagrams (b),
illustrated here for the $B_d^0-\overline{B}_d^0$ case.}
\end{figure}
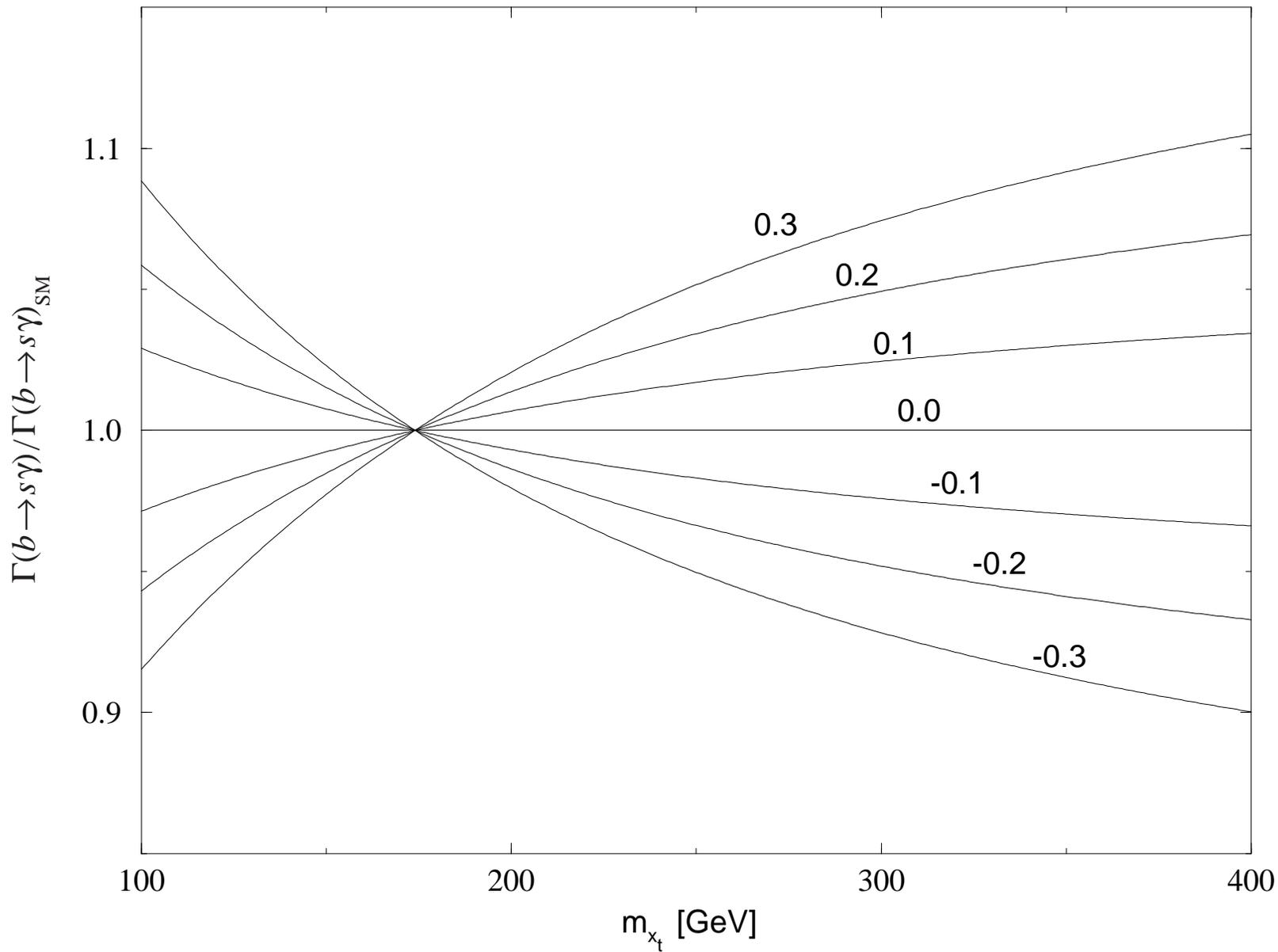
\begin{figure}
\caption{ \label{fig:bsgam}
Effects of a $Q={2\over 3}$ singlet quark on $b\to s\gamma$ decay.
The branching fraction normalized to the SM value is shown versus the
singlet quark mass. The curves are labelled by the values of
$|\hat V_{xs}^*\hat V^{}_{xb}|$; we assume that
$\hat V_{xs}^*\hat V^{}_{xb}$ and $\hat V_{ts}^*\hat V^{}_{tb}$
have the same phase within an overall $\pm$ sign, shown on the
label.}
\end{figure}
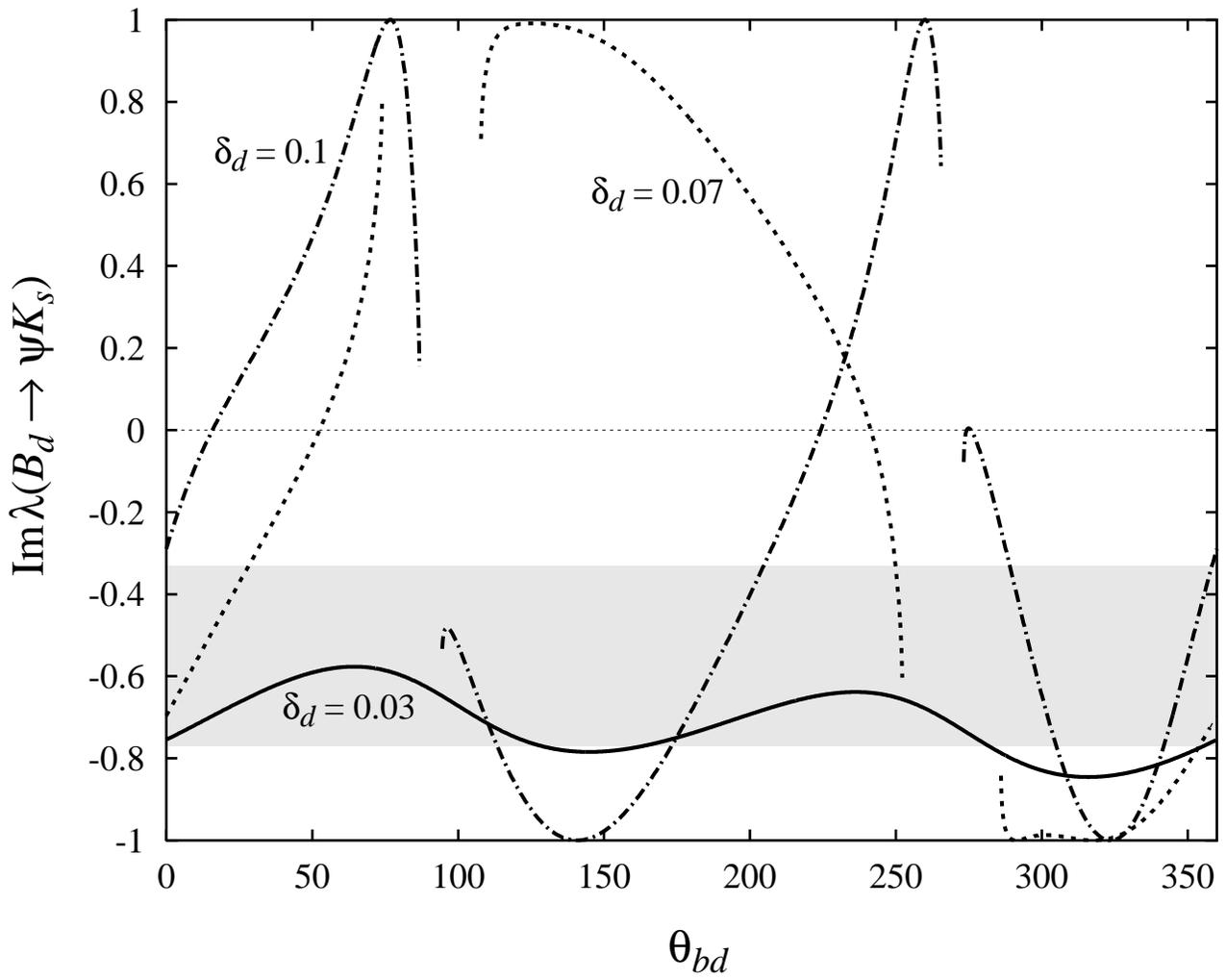
\begin{figure}
\caption{ \label{fig:beta} The CP asymmetry
${\rm Im } \; \lambda (B_d\to \psi K_S)$ in the presence
of down-type singlet quarks. The band indicates the present uncertainty
in the SM prediction. For some values of $\delta _d$ and $\theta _{bd}$
(defined by Eq.~(\protect\ref{eq:defin})) there are no solutions as the
unitarity quadrangle cannot be made to close. }
\end{figure}
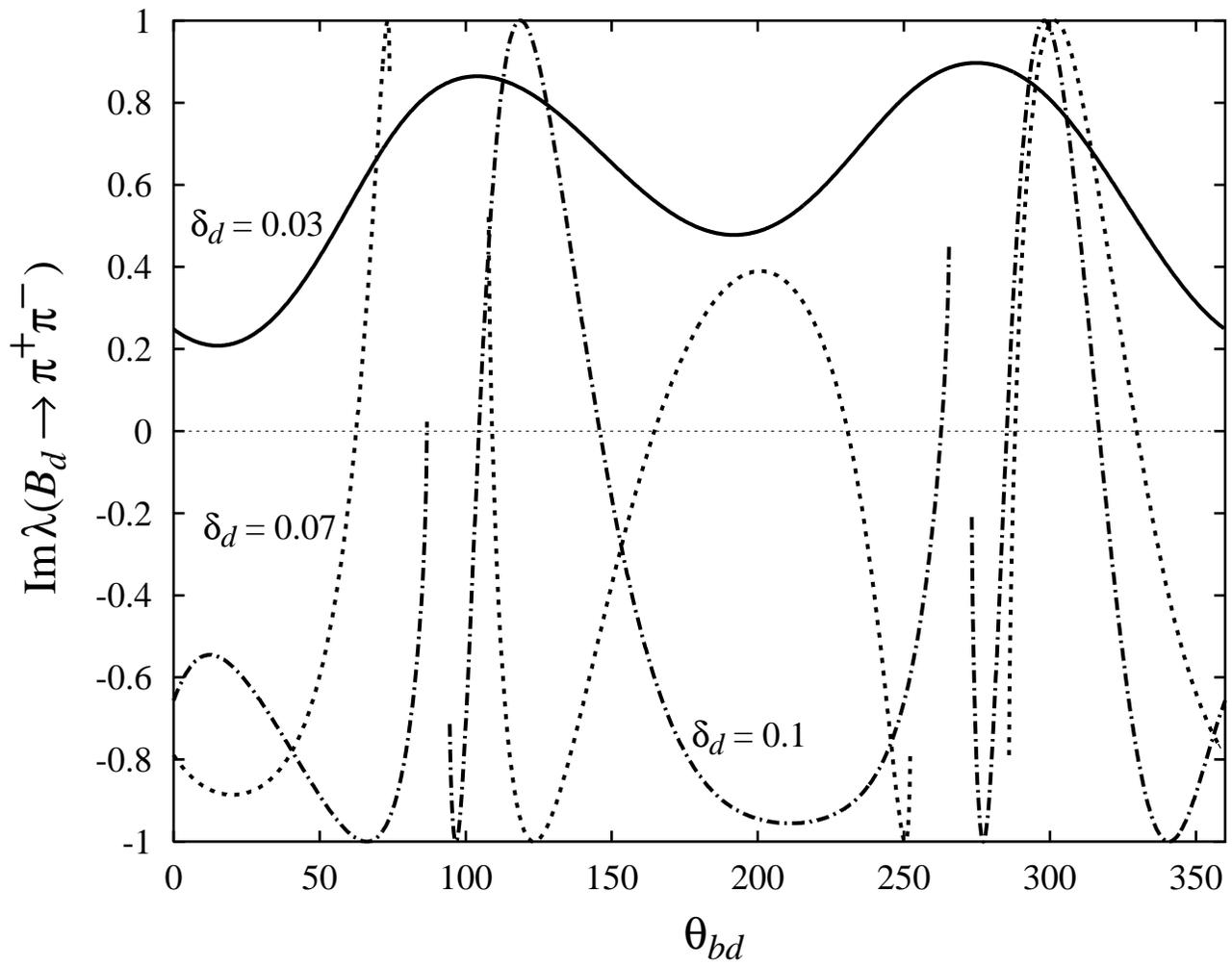
\begin{figure}
\caption{ \label{fig:alpha} The CP asymmetry
${ \rm Im }\; \lambda (B_d\to \pi ^+\pi ^-)$
in the presence of down-type singlet quarks. The entire range is allowed
in the SM.  The parameters $\delta _d$ and $\theta _{bd}$ are defined by
Eq.~(\protect\ref{eq:defin}).}
\end{figure}
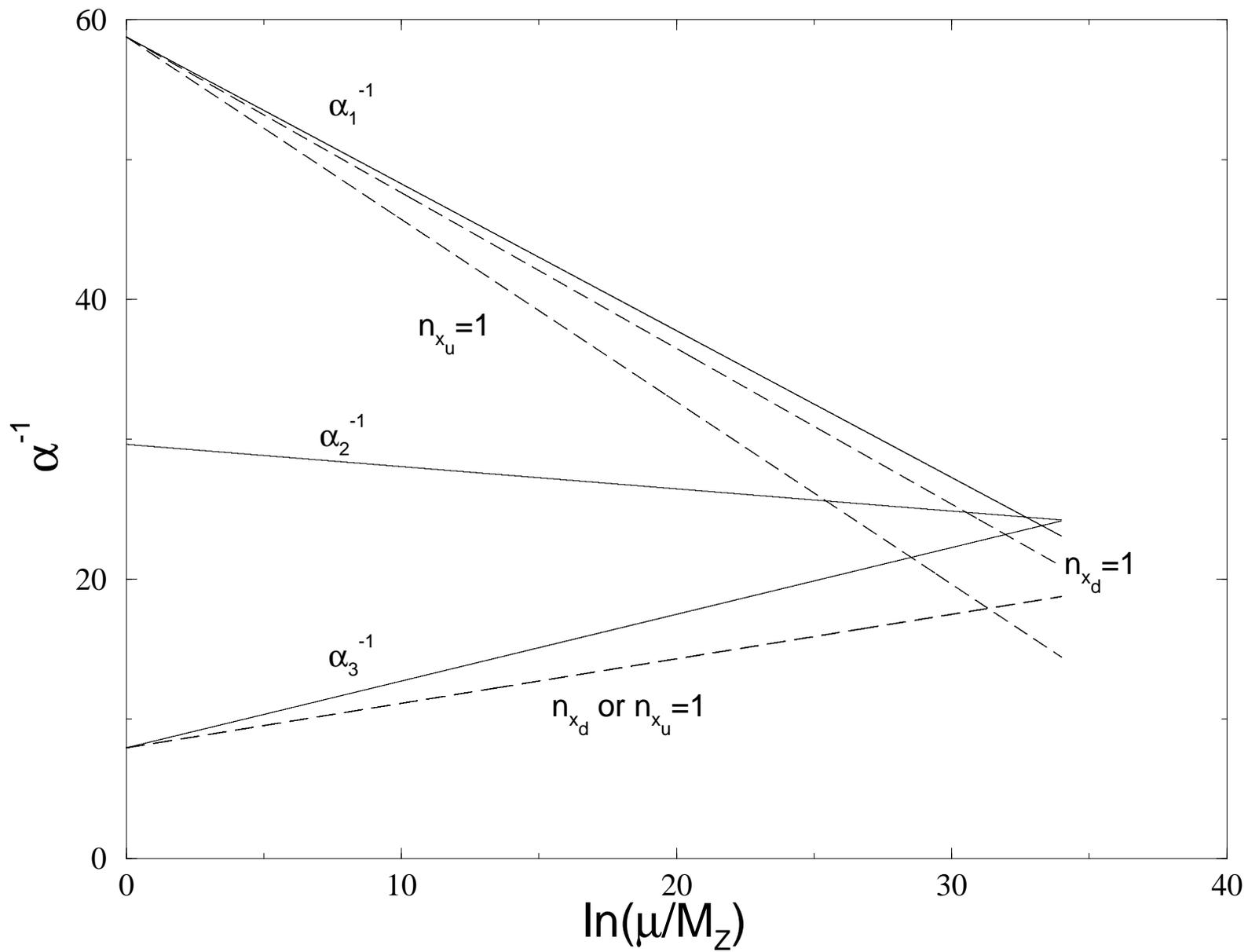
\begin{figure}
\caption{ \label{fig:singlet} One-loop gauge coupling evolution,
adding either one $d$-type singlet ($n_{x_d}=1)$ or one $u$-type
singlet ($n_{x_u}=1$) to the MSSM. The singlets leave the $SU(2)_L$ gauge
coupling unaffected at one-loop, but alter the evolution of the other
gauge couplings and destroy unification.}
\end{figure}
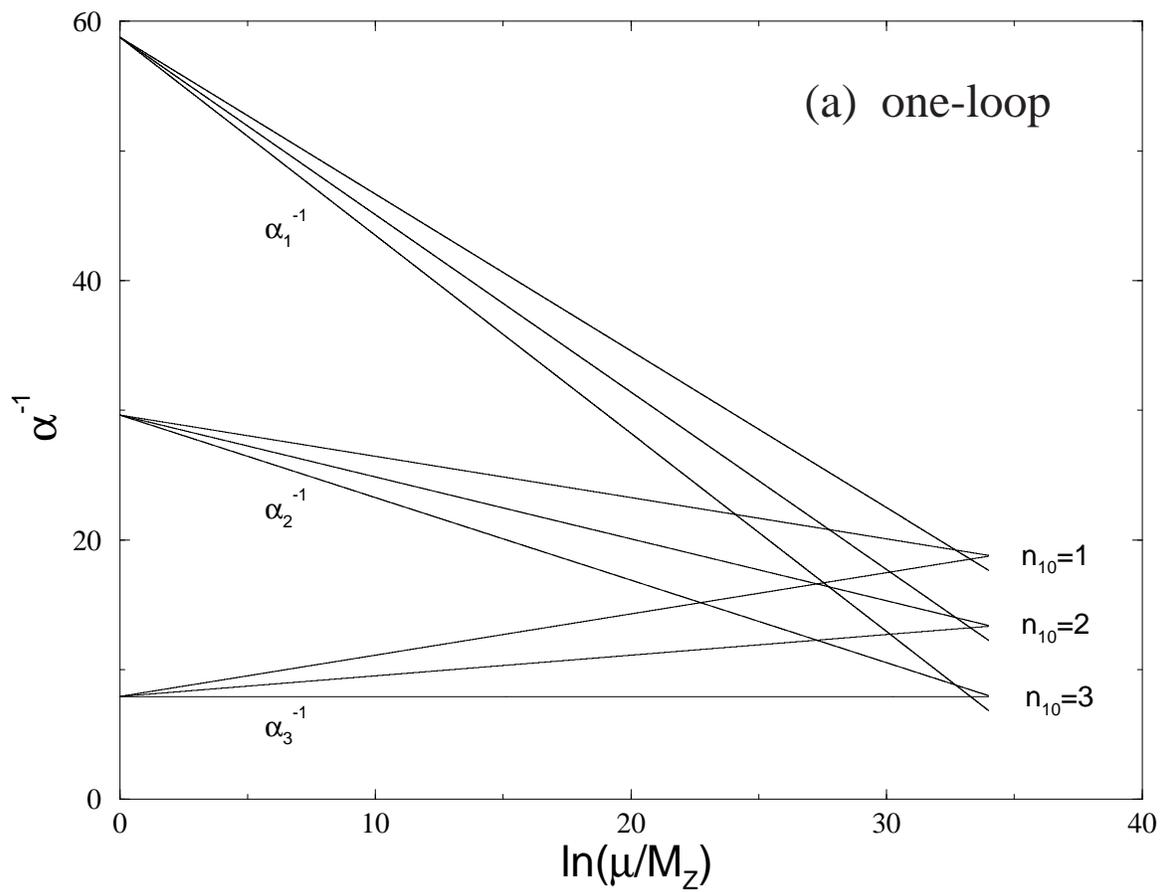
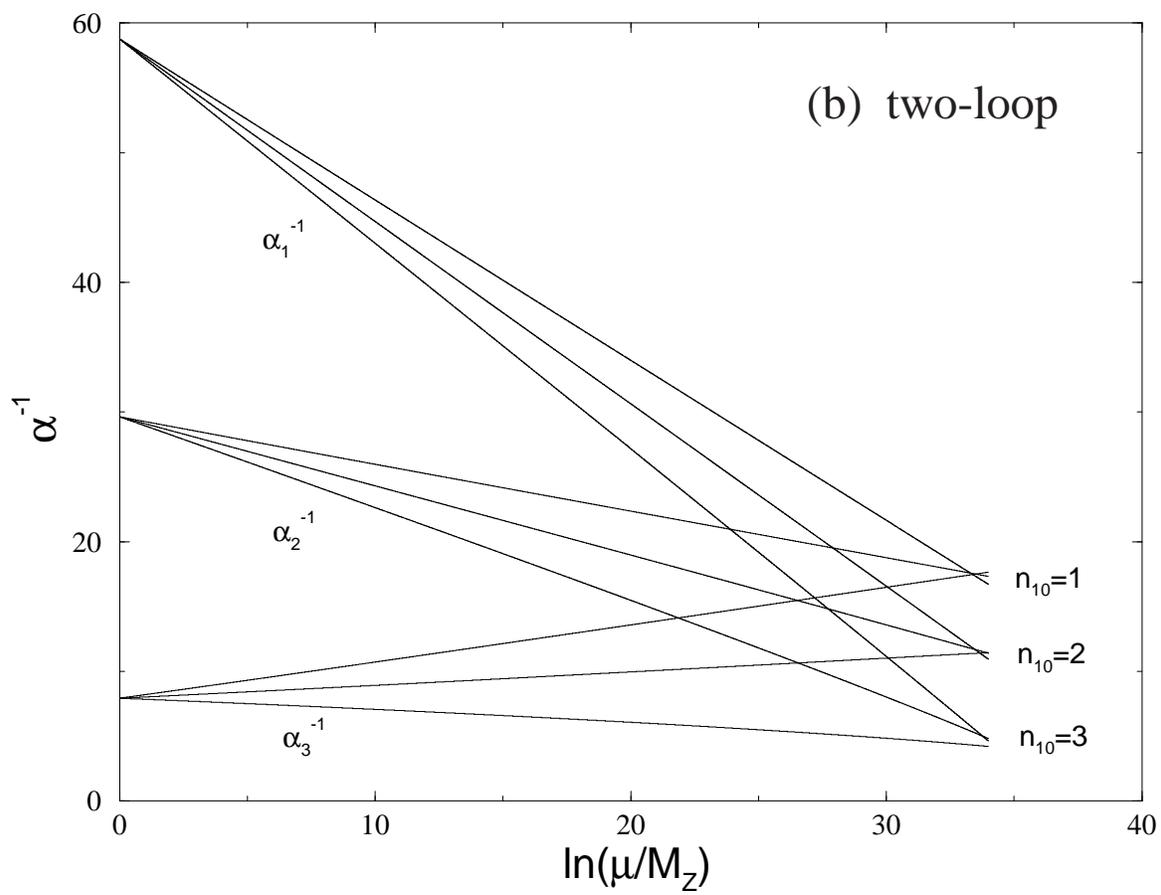
\begin{figure}
\caption{ \label{fig:oneg1h} One-loop and two-loop gauge coupling evolution
with the addition of different numbers of light ${\bf 10}$ multiplets of
$SO(10)$ to the MSSM. Successful gauge coupling unification
is preserved with the addition of one or two ${\bf 10}$-plets, but
is threatened by two-loop effects when three light ${\bf 10}$-plets
are added to the MSSM. }
\end{figure}
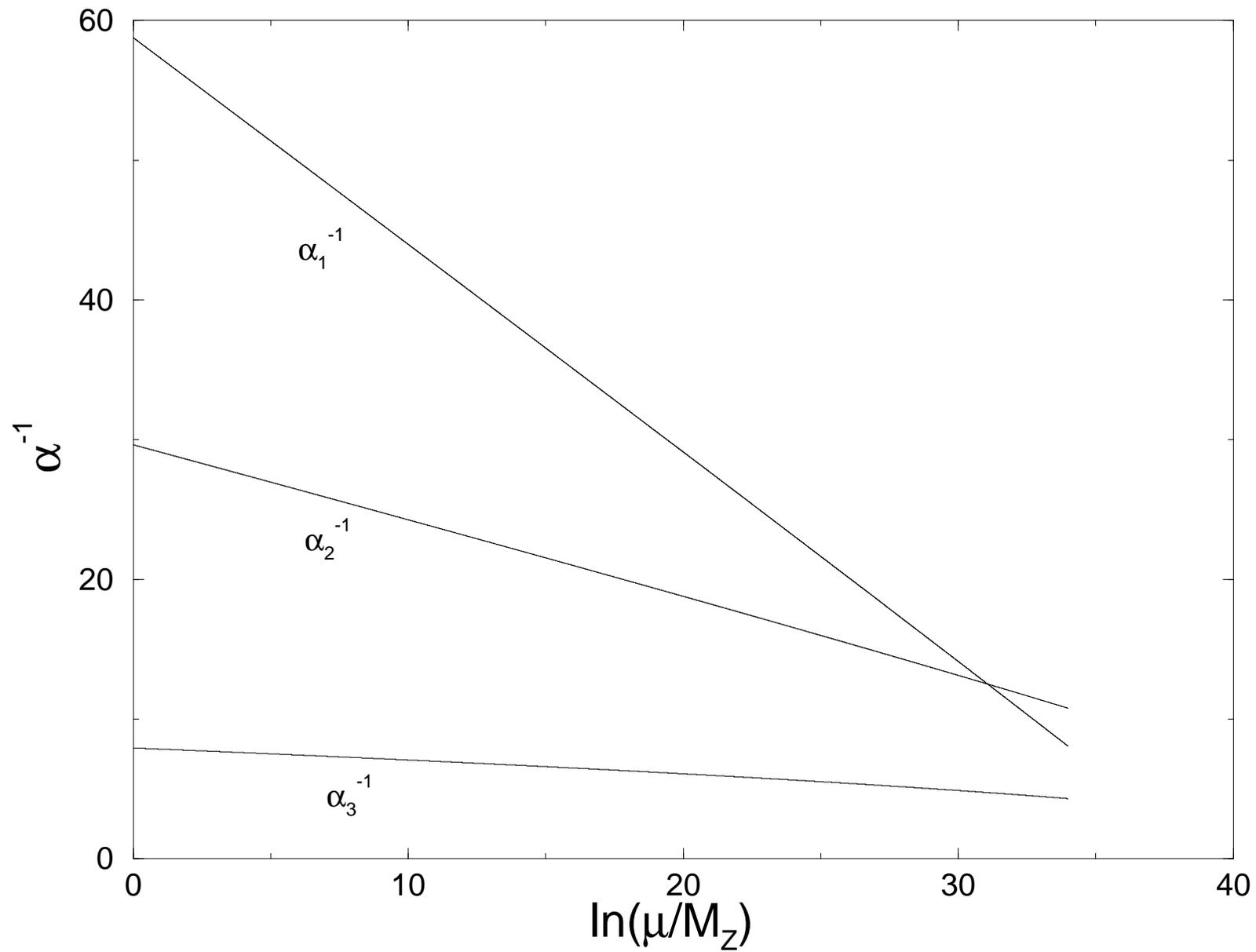
\begin{figure}
\caption{\label{fig:threeg} Two-loop gauge coupling evolution where the
electroweak scale matter consists of three light ${\bf 27}$'s.
Gauge coupling unification is unsuccessful
without the presence of a light pair of Higgs doublets
from a split representation. The result for the case including a pair of
light Higgs doublets is shown in Fig.~(\protect\ref{fig:oneg1h}).}
\end{figure}
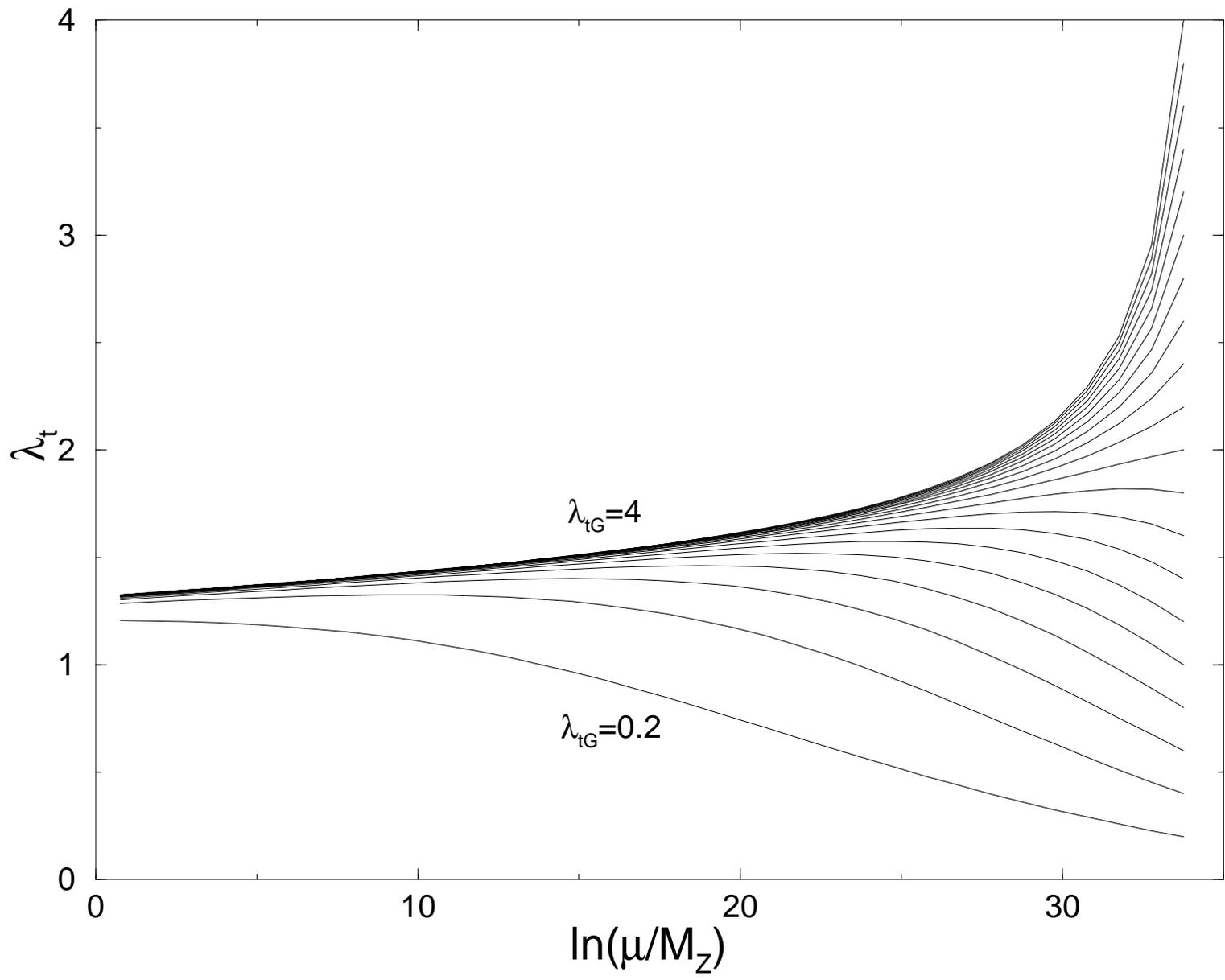
\begin{figure}
\caption{\label{fig:yukawa} The evolution of the top quark Yukawa coupling
in the presence of three light ${\bf 10}$ multiplets of $SO(10)$ added to
the MSSM. The top quark Yukawa coupling reaches its infrared
fixed point for a large range of initial (GUT) values, $\lambda _{tG}$.
The curves show $\lambda _{tG}$ between 0.2 and 4 in increments of 0.2.}
\end{figure}
\end{document}